\documentclass[a4paper,11pt]{article}

\usepackage{geometry}
\usepackage{authblk}
\usepackage{hyperref}       
\usepackage[backend=biber,style=authoryear]{biblatex}
\usepackage{graphicx}%
\usepackage{amsmath,amssymb,amsfonts,amsthm}%
\usepackage{mathrsfs}%
\usepackage[title]{appendix}%
\usepackage{xcolor}%
\usepackage{textcomp}%
\usepackage{manyfoot}%
\usepackage{booktabs}%
\usepackage{listings}%
\usepackage{url}%

\usepackage{xspace}
\usepackage{cleveref}   
\usepackage{mathtools}
\DeclarePairedDelimiter{\norm}{\lVert}{\rVert}
\usepackage{tabularx}
\usepackage{rotating}
\usepackage{pdflscape}
\usepackage{csquotes}
\usepackage{multirow}   
\usepackage[font={bf}, tableposition=top]{caption} 
\usepackage{subcaption} 
\usepackage{bold-extra} 
\usepackage[vlined,linesnumbered,ruled,noend]{algorithm2e} 
\usepackage{microtype}  
\usepackage{siunitx}    
\usepackage{xfrac}      
\usepackage{physics}
\usepackage{dsfont}
\usepackage{soul}

\newcommand{\spara}[1]{\smallskip\noindent\textbf{#1}}

\newenvironment{squishlist}
{\begin{list}{$\bullet$}
 {\setlength{\itemsep}{0pt}
     \setlength{\parsep}{3pt}
     \setlength{\topsep}{3pt}
     \setlength{\partopsep}{0pt}
     \setlength{\leftmargin}{1.5em}
     \setlength{\labelwidth}{1em}
     \setlength{\labelsep}{0.5em} } }
{\end{list}}

\usepackage{xcolor}
\graphicspath{{fig}}
\definecolor{verylightgray}{rgb}{0.8,0.8,0.8}
\definecolor{verylightgray}{rgb}{0.8,0.8,0.8}
\definecolor{brown}{rgb}{0.55, 0.25, 0.0}
\definecolor{forestgreen}{rgb}{0.0, 0.5, 0.0}
\definecolor{blue-violet}{rgb}{0.54, 0.17, 0.89}
\definecolor{dartmouthgreen}{rgb}{0.05, 0.5, 0.0}
\definecolor{dark-red}{rgb}{0.8, 0.0, 0.0}
\definecolor{light-blue}{rgb}{0.5, 0.5, 0.99}
\definecolor{brinkpink}{rgb}{0.85, 0.25, 0.5}
\definecolor{columbiablue}{rgb}{0.61, 0.87, 1.0}
\definecolor{cyan(process)}{rgb}{0.0, 0.55, 0.85}
\definecolor{darkcyan}{rgb}{0.0, 0.0, 0.8}
\definecolor{darkorange}{rgb}{0.85, 0.45, 0.0}
\definecolor{deeplilac}{rgb}{0.6, 0.33, 0.73}
\definecolor{electricultramarine}{rgb}{0.25, 0.0, 1.0}
\definecolor{electricviolet}{rgb}{0.56, 0.0, 1.0}

\theoremstyle{plain}
\newtheorem{theorem}{Theorem}

\theoremstyle{plain}
\newtheorem{definition}[theorem]{Definition}

\newtheorem{impdefinition}{Implementation of Definition}
\theoremstyle{plain}

\theoremstyle{plain}

\theoremstyle{plain}
\newtheorem{problem}{Problem}
\newtheorem{impproblem}{Implementation of Problem}

\newcolumntype{C}[1]{>{\centering\arraybackslash}m{#1}}

\DeclareMathOperator*{\argmin}{arg\,min}
\newcommand{\at}[2][]{#1|_{#2}}
\newcommand{\val}{\ensuremath{v}\xspace}
\newcommand{\pl}{\ensuremath{\mathrm{P\&L}}\xspace}
\newcommand{\historicalpl}{\ensuremath{r}\xspace}
\newcommand{\historicalpls}{\ensuremath{\mathbf{r}}\xspace}
\newcommand{\Vega}{\ensuremath{\mathcal{V}}\xspace}
\newcommand{\ui}{\ensuremath{G}\xspace}
\newcommand{\uei}{\ensuremath{H}\xspace}
\newcommand{\asstP}{\ensuremath{\mathcal{P}}\xspace}
\newcommand{\asstG}{\ensuremath{\mathcal{G}}\xspace}
\newcommand{\asstEOS}{\ensuremath{\mathcal{H}}\xspace}
\newcommand{\asstTotal}{\ensuremath{\mathcal{T}}\xspace}
\newcommand{\notionalP}{\ensuremath{\mathbf{p}}\xspace}
\newcommand{\notionalG}{\ensuremath{\mathbf{g}}\xspace}
\newcommand{\notionalEOS}{\ensuremath{\mathbf{h}}\xspace}
\newcommand{\notionalTotal}{\ensuremath{\mathbf{t}}\xspace}

\newcommand{\uiuniverse}{\ensuremath{\mathcal{U}}\xspace}
\newcommand{\domain}{\ensuremath{\mathcal{D}}\xspace}
\newcommand{\UEIsuniverse}{\ensuremath{\uiuniverse^\asstEOS}\xspace}
\newcommand{\initialportfolio}{\ensuremath{\langle\asstP, \notionalP \rangle}\xspace}
\newcommand{\eligibleoptimizationstrategy}{\ensuremath{\langle\asstEOS, \notionalEOS \rangle}\xspace}
\newcommand{\totalportfolio}{\ensuremath{\langle\asstTotal, \notionalTotal\rangle}\xspace}
\newcommand{\generalportfolio}{\ensuremath{\langle\asstG, \notionalG\rangle}\xspace}
\newcommand{\optimaltotalportfolio}{\ensuremath{\langle\asstTotal^\star, \notionalTotal^\star\rangle}\xspace}
\newcommand{\estimatedoptimaltotalportfolio}{\ensuremath{\langle \widehat{\asstTotal}, \widehat{\notionalTotal} \rangle}\xspace}
\newcommand{\expectedpl}{\ensuremath{\overline{\mathrm{P\&L}}}\xspace}
\newcommand{\expectedplT}{\ensuremath{\overline{\mathrm{P\&L}}^{\mathcal{T}}}\xspace}
\newcommand{\varr}{\ensuremath{\beta\text{-}\mathrm{VaR}}\xspace}
\newcommand{\varrT}{\ensuremath{\beta\text{-}\mathrm{VaR}^{\mathcal{T}}}\xspace}

\newcommand{\sort}[1]{\ensuremath{\mathcal{S}\left(#1\right)}\xspace}
\newcommand{\nat}{\ensuremath{\mathbb{N}}\xspace}
\newcommand{\integer}{\ensuremath{\mathbb{Z}}\xspace}
\newcommand{\reall}{\ensuremath{\mathbb{R}}\xspace}
\newcommand{\EOSset}{\ensuremath{\mathcal{E}}\xspace}

\newcommand{\algorithmname}{RATS\xspace}

\newcommand{\x}{\ensuremath{\mathbf{x}}\xspace}

\newcommand{\myv}{\ensuremath{\mathbf{v}}\xspace}

\newcommand{\cpers}{\ensuremath{c^{\mathrm{pers}}\xspace}}
\newcommand{\csoc}{\ensuremath{c^{\mathrm{soc}}\xspace}}
\newcommand{\rone}{\ensuremath{\mathbf{r}_1}\xspace}
\newcommand{\rtwo}{\ensuremath{\mathbf{r}_2}\xspace}
\newcommand{\pbest}{\ensuremath{\mathbf{y}}\xspace}
\newcommand{\gbest}{\ensuremath{\mathbf{z}}\xspace}

\begin{document}

\title{Risk-aware Trading Portfolio Optimization}

\author[1,2,\dag]{Marco~Bianchetti}
\author[3]{Gabriele~D'Acunto\thanks{The work of Gabriele D'Acunto was supported by CENTAI Institute, Turin, Italy, and conducted during his affiliation with the Institute.}}
\author[4]{Gianmarco~De~Francisci~Morales}
\author[4]{Yuko~Kuroki}
\author[5]{Marco~Scaringi\thanks{The present study was conducted while Marco Scaringi was affiliated with Financial \& Market Risk Management, Intesa Sanpaolo, Milan, Italy.}}
\author[4]{Fabio~Vitale}

\affil[1]{Financial \& Market Risk Management, Intesa Sanpaolo, Milan, Italy}
\affil[2]{Department of Statistical Sciences \enquote{Paolo Fortunati}, University of Bologna, Italy}
\affil[3]{Department of Information Engineering, Electronics and Telecommunications, Sapienza University, Rome, Italy}
\affil[4]{CENTAI, Turin, Italy}
\affil[5]{Risk Trading Quant Department, ING Bank N. V., Milan, Italy}
\affil[$\dag$]{Corresponding author, \texttt{marco.bianchetti@unibo.it}}

\date{}

\maketitle 

\begin{abstract}
We investigate portfolio optimization in financial markets from a trading and risk management perspective. 
We term this task Risk-Aware Trading Portfolio Optimization (RATPO), formulate the corresponding optimization problem, and propose an efficient Risk-Aware Trading Swarm (\algorithmname) algorithm to solve it.
The key elements of RATPO are a generic initial portfolio $\mathcal{P}$, a specific set of Unique Eligible Instruments (UEIs), their combination into an Eligible Optimization Strategy (EOS), an objective function, and a set of constraints. \algorithmname searches for an optimal EOS that, added to $\mathcal{P}$, improves the objective function repecting the constraints. 

\algorithmname is a specialized Particle Swarm Optimization method that leverages the parameterization of $\mathcal{P}$ in terms of UEIs, enables parallel computation with a large number of particles, and is fully general with respect to specific choices of the key elements, which can be customized to encode financial knowledge and needs of traders and risk managers.

We showcase two RATPO applications involving a real trading portfolio made of hundreds of different financial instruments, an objective function combining both market risk (VaR) and profit\&loss measures, constrains on market sensitivities and UEIs trading costs.
In the case of small-sized EOS, \algorithmname successfully identifies the optimal solution and demonstrates robustness with respect to hyper-parameters tuning. 
In the case of large-sized EOS, \algorithmname markedly improves the portfolio objective value, optimizing risk and capital charge while respecting risk limits and preserving expected profits.

Our work bridges the gap between the implementation of effective trading strategies and compliance with stringent regulatory and economic capital requirements, allowing a better alignment of business and risk management objectives.
\end{abstract}

\newpage

\tableofcontents

\vspace{2cm} \noindent 

\vspace{0.5cm}\noindent 

\vspace{0.5cm}\noindent 

\vspace{0.5cm}\noindent 
\textbf{Acknowledgements}: the authors contributed equally to this work. The authors acknowledge fruitful discussions with many colleagues from Intesa Sanpaolo Risk Management and Front Office Departments, Francesco Terribile and Pietro Gallo from  Politecnico di Milano, who collaborated to the early stage of this work, Luca Lamorte, who joined at later stage, and Laura Li Puma, Arianna Miola, and Luigi Ruggerone from Intesa Sanpaolo Innovation Center for supporting the research team. 
Marco Scaringi declares that the views and opinions expressed here are his own and do not represent the views of his previous or current employers; they are not responsible for any use that may be made of these contents.

\newpage

\section{Introduction}
\label{sec:introduction}

The financial crisis started in 2007, and the subsequent period of severe market stresses revealed weaknesses in the risk capital framework of the global banking and financial system. Since then, regulators increased pressure towards risk-based capital reserves.
The Basel Committee on Banking Supervision (BCBS), the main international body that develops global banking standards, issued the so-called Basel III and Basel IV accords (\cite{Basel3,Basel4}), introducing substantial amendments to the capital treatment of all risk classes (market, liquidity, counterparty, credit and operational risk).\footnote{The BCBS also maintains the ``Basel Framework" (\url{https://www.bis.org/basel_framework}) which provides a comprehensive and updated compendium of BCBS financial regulations.}
The BCBS proposals are transposed into legal frameworks by national and international authorities.\footnote{The European Union transposed the after-crisis BCBS proposals into the so-called CRD4/CRR package in 2013 (\cite{CRD4,CRR1}), the amended CRD5/CRR2 package in 2019 (\cite{CRD5,CRR2}), and the further amended CRD6/CRR3 package in 2024 (\cite{CRD6,CRR3}), the latter to enter into force in January 2025 and January 2026, respectively. The European Banking Authority (EBA)  maintains the Interactive Single Rulebook (\url{https://www.eba.europa.eu/single-rulebook}) which provides a comprehensive compendium of EU financial regulations and related Q\&As.}

The trading and risk management practices of financial institutions has developed according to regulatory requests. Nowadays, banks typically put huge efforts into research, development, implementation, and production of several different risk measures, which are the main inputs to compute capital reserves, both for regulatory purposes (regulatory capital and Risk-Weighted Assets---RWA) and internal managerial purposes (economic capital). 
In particular, market risk measures (e.g., Value at Risk, Stressed Value at Risk, Incremental Risk Charge) are typically computed with daily frequency, both for managerial reporting to trading desks, and to determine quarterly averages for regulatory reporting of capital reserves. Budgets for human and IT resources for risk management purposes have increased accordingly.

\paragraph{Our goal: risk-aware trading portfolio optimization}
In the \enquote{classic} portfolio optimization approach \`{a} la Markowitz (\cite{Mar1952}), the point of view of portfolio managers typically prevails, who focus on \emph{dynamic} forward-looking estimates of return and risk by selecting a portfolio of assets and looking for a strategy to adjusts the assets' weights over a given time horizon $T-t$, where $t$ is the rebalancing date.

In this work, we focus on a different problem, named \emph{risk-aware trading portfolio optimization}, where we take the point of view of the bank's traders and risk managers, who focus on \emph{statical} portfolio risk measurement and features at a given point in time $t$, and look for optimal trades at the same time $t$ to reduce capital reserves while preserving portfolio value, consistently with the business objectives. This view, characterizing our approach, represents a stark difference with Markowitz's framework.

Another distinguishing feature of our approach concerns the full generality with respect to the portfolio composition.
More in detail, we consider \emph{large heterogeneous trading portfolios} typical of commercial or investment banks, potentially encompassing millions of financial instruments such as stocks, securities, loans, derivatives, and funds, depending on any underlying \emph{risk factors}, i.e., interest rates, credit, equity, forex, and commodities. 
These positions can feature any level of \emph{complexity}: linear and non-linear payoffs, plain vanilla and exotic options, multi-asset and hybrid derivatives. 
They are managed through appropriate pricing models and computational algorithms such as analytical formulas and Monte Carlo simulations.
Moreover, trading portfolios are characterized by continuous \emph{turnover}, driven by expiry of old trades, flows of new trades with clients, proprietary investments, speculative or hedging strategies, and generate continuous in/out cash flows related to received/paid coupons, collateral exchange, financing operations, and fees.

Furthermore, in our approach portfolio risks are measured through various \emph{risk measures} that capture all the different risk sources, e.g., market risk, liquidity risk, credit and counterparty risk, and valuation risk.\footnote{Valuation risk is captured by appropriate \emph{valuation risk measures} such as fair value adjustments, required by EU accounting standards (IFRS, \url{https://www.ifrs.org}), and Additional Valuation Adjustments (AVA) deducted from Common Equity Tier 1 capital, required by EU prudential regulation (\cite{CRR3}), etc., which ultimately determine the capital structure of the Bank.}
These risks are managed within a global Risk Appetite Framework (RAF)\footnote{RAF guidelines are issued by the Basel Committee of Banking Supervision (\cite{BCBS15Corporate}) and transposed by the EU into CRD/CRR packages.}, a structured approach used by Banks to define, manage, and align their willingness to assume risks in pursuit of their business objectives. The RAF is based on \emph{risk limits}, which are set, monitored, and reported by a risk management unit independent of business units.

\paragraph{Core components} 
Several crucial components characterize our risk-aware trading portfolio optimization approach.
\emph{unique eligible instruments} are the main building blocks used to optimize the portfolio: they uniquely identify liquid market instruments quoted by exchanges, brokers, or market makers, typically used by traders for risk management purposes, i.e., to hedge Delta, Vega, and Gamma risks.
Portfolio optimization is encoded through \emph{eligible optimization strategies}, i.e., combinations of unique eligible instruments added to the portfolio to optimize the selected \emph{objective function}, while complying with a set of \emph{constraints} and adhering to a specific \emph{structure} reflecting the prior financial knowledge of the bank's traders and risk managers. 
The objective function consists of a combination of portfolio risk measures and features such as profit and loss measures and market transaction costs.
Instead, the constraints are associated with risk limits to effectively drive the optimization process.
More details and practical examples of these core components are included in the following sections.

\paragraph{Challenges}
Our risk-aware trading portfolio optimization problem presents numerous challenges. The objective function is typically non-linear and its computation, particularly in the case of risk measures based on historical or Monte Carlo simulation, can require substantial computational resources. 
The unique eligible instruments are determined by various parameters, which are either intrinsically categorical (e.g., payoff type, currency, underlying asset), or are quoted in market-discrete values (e.g., maturity, strike), though some can be approximated continuously (e.g., notional amounts). 
Hence, a candidate eligible optimization strategy is represented by a set of values of the relevant parameters.  
 
However, the sheer number of potential combinations of the parameters' values makes the brute-force approach (i.e., the direct calculation of all possible solutions) unfeasible even for simple portfolios and optimization strategies. 
Additionally, not every optimization strategy satisfies the optimization constraints, depending on their tightness (the tighter the constraints, the fewer eligible solutions).
Finally, the global optimal solution is often hidden among numerous similar local optima due to the inherent complexity of the objective function landscape.

\paragraph{Optimization meta-heuristics}
From a computational point of view, we must solve a multidimensional, non-linear, constrained optimization problem with integer variables, where the value of the objective function may change abruptly, thus complicating the search for a global minimum.
Since simple techniques are insufficient for such complex problems, numerical algorithms based on \emph{optimization meta-heuristics} are typically used.
These techniques are recognized as efficient approaches for optimization problems that cannot be exactly solved in a “reasonable” time limit, overcoming incomplete initial information or limited computation capacity, and sampling a portion of the entire set of solutions that is too large to be completely explored. 
Even though these algorithms do not guarantee to converge to the global optimum in a finite computational time, their application to our optimization problem is beneficial, since also local optima leading to appreciable improvements of the objective function are valuable from a financial point of view.

A complete overview of optimization meta-heuristics is presented by \cite{BouLep13}, who distinguish between \emph{single solution based} algorithms, such as Simulated Annealing (\cite{Kirk1983}), which start with a single initial solution and move along a trajectory in the search space, and \emph{population based} algorithms, which explore the search space using a population of different solutions evolving according to some rules, such as Genetic Algorithms (GA, \cite{Hol1992}), and Particle Swarm Optimization (PSO, \cite{KE95}).

\paragraph{Related literature}
The classic portfolio optimization problem following \cite{Mar1952} is widely discussed in the literature, also using optimization meta-heuristics (\cite{Dallagnol2009,Soleimani2009}) and even quantum computers (\cite{Rosenberg16,VenKon19}). 
Conversely, to the best of our knowledge, the risk-aware trading portfolio optimization problem introduced above represents an understudied topic.
Our work was initially motivated by \cite{KonGeo17b}, who use GA and PSO meta-heuristics to search optimal delta hedging strategies in the context of Margin Valuation Adjustment optimization.\footnote{Margin Valuation Adjustment is a valuation adjustment which takes into account the cost of financing initial margin collateralization of OTC derivatives, see e.g. \cite{Greg20,Green15}.}
This seminal work is limited to a specific and very simple setting: a toy portfolio of USDCNY Non Deliverable FX Forwards (NDF, i.e., linear instruments on a single exchange rate) exchanged between two out of five counterparties, one single optimization instrument (the USDCNY NDF itself), and no optimization constraints beyond trading cost.
An apparent application to collateral optimization with multiple Central Counterparties is found in \cite{Ere17}.

The idea of portfolio optimization based on market risk measures is obviously not new. 
For example, \cite{RocUry00} and \cite{Larsen2002} minimize Conditional Value at Risk (a.k.a. Expected Shortfall)  using analytical approaches and linear programming techniques. 
\cite{Gaivoronski2005} determine an optimal Value at Risk portfolio with a given expected return.
\cite{Gilli2006} determine optimal quantities of $n$ assets which minimize different risk functions for a given return target under constraints on the assets' sizes.
\cite{Faias2011}, \cite{Faias2017} and \cite{Dao14} consider simplified European options' portfolios.
  
Also the idea of portfolio optimization based on meta-heuristics is largely explored in the literature. \cite{Chang2009} solve portfolio optimization problems by employing GA and risk measures based on Markowitz's mean-variance, semi-variance, mean absolute deviation, and variance with skewness. 
\cite{Deng2012} introduce an improved PSO to solve the Cardinality Constraints Markowitz Portfolio Optimization problem.
\cite{Das2023} compare PSO, GA, Dynamic Programming, and Differential Evolutionary Algorithm algorithms for portfolio optimization in the NIFTY 50 market using Sharpe Ratio and expected return.
\cite{JunJohar2023} use PSO to maximize the Sharpe Ratio of portfolios of Exchange-Traded Funds, demonstrating its effectiveness in achieving higher risk-adjusted returns compared to traditional methods. 
\cite{Erwin2023} develop a Multi-Guide Set-Based Particle Swarm Optimization for multi-objective portfolio optimization, and test it against established algorithms. 
 
Excellent reviews by \cite{Fabozzi2010}, \cite{Ertenlice2018}, \cite{Erwin2023meta}, \cite{Loke2023} and \cite{Gunjan2023} are available.
\cite{Erwin2023meta} review 140 studies using evolutionary and swarm intelligence algorithms for portfolio optimization, finding that meta-heuristic approaches as more computationally efficient.

\paragraph{Our contributions}
With the partial exception of \cite{KonGeo17b}, none of the cited works addresses the challenges of the risk-based trading portfolio optimization problem investigated in this paper. 
Specifically, our contributions are outlined below.
\begin{squishlist}
    \item We introduce and formalize the risk-aware trading portfolio optimization problem, taking the point of view of the bank’s traders and risk managers.
    Toward our goal, we introduce and consequently leverage the key concepts of the unique eligible instruments and eligible optimization strategies, also including options to allow Delta, Vega, and Gamma optimization.   
    \item We propose a modified PSO algorithm, called \emph{Risk-Aware Trading Swarm} (\algorithmname), incorporating a specialized parameterization built around the concepts of the unique eligible instruments and eligible optimization strategy. 
    \algorithmname is a versatile framework that leverages the parallelizable nature of the optimization problem with respect to the particle dimension and can be adapted to different trading portfolios, objective functions, and constraints.
    Additionally, its parameterization enables users to incorporate their financial insights and business views by directly shaping the structure of the eligible optimization strategies.
    \item We showcase two applications on a real trading portfolio typical of large banks, made of hundreds of financial instruments of different kinds, using real market data and tradable optimization instruments.
    In both cases, we seek an eligible optimization strategy that, when combined with the initial portfolio, optimizes a fractional, non-convex objective function over a non-convex set of eligible optimization strategies defined by the constraints. 
    Inspired by the practical needs of traders and risk managers, the objective function accounts for the Value at Risk measure, expected P\&L, and trading costs. In addition, the set of constraints includes sensitivity limits; while the size of the eligible optimization strategy, its diversification, and the maximum nominal amount are handled directly via the proposed parameterization.
    In the first application, focused on a small-sized optimization strategy, \algorithmname successfully identifies an optimal solution within the optimal solution set and demonstrates robustness to hyper-parameters tuning across all tested scenarios. 
    In the second application, involving a large-sized optimization strategy, \algorithmname markedly improves the cost-adjusted expected P\&L and risk measures of the initial portfolio in every setting.  
\end{squishlist}
\medskip
At a high level, our work overcomes our inspiring paper by \cite{KonGeo17b} and bridges the gap between complying with stringent regulatory requirements and implementing effective trading strategies that align with the objectives of both business units and risk management.

\paragraph{Roadmap} 
This paper is organized as follows.
\Cref{sec:preliminaries} provides the fundamental concepts underlying the risk-aware trading portfolio problem.
Then, \Cref{sec:ratpo} formalizes the risk-aware trading portfolio optimization problem.
Next, \Cref{sec:RATS} describes the proposed method, i.e., \algorithmname.
\Cref{sec:empirical_assessment} provides the results of our empirical assessment. In detail, \Cref{subsec:expset} specifies the setting and the building blocks of the risk-aware trading portfolio optimization problem.
\Cref{subsec:M_approach} compares our approach to Markowitz-like approaches, highlighting their limitations and key differences.
Subsequently, \Cref{subsec:toy_portfolio} explores the small-sized application setting, and \Cref{subsec:real_portfolio} addresses the large-sized one. Finally \Cref{sec:conclusions} draws the conclusions and outlines future research directions.

\section{Preliminaries}\label{sec:preliminaries}
This section provides the basic notation and key concepts that underlie the risk-aware trading portfolio problem.
We start the discussion at the individual financial instrument level and then move to the portfolio level.

\paragraph{Notation}
We denote by $[n]$ the set $\{1,2,\ldots,n\}$ with $n\in\mathbb{N}$. 
Scalars are lowercase letters, $a$, vectors are lowercase bold, $\mathbf{a}$, matrices are uppercase bold, $\mathbf{A}$, sets are uppercase calligraphic, $\mathcal{A}$, elements of a set are uppercase, $A$.
Given $a$, we denote $\max(a,0)$ by $[a]_+$.
The multiset union operator is $\uplus$.
The element-wise (Hadamard) product is denoted by $\circ$.
We use subscripts for indices, while superscripts are used for everything else.
Given an $n$-dimensional vector $\mathbf{a}$ and a permutation $\prec$ such that $\mathbf{a}^\prec=[a^\prec_1, \dots, a^\prec_n]$ is increasing, we denote by $\sort{\cdot}$ the sort operator such that $\mathbf{a}^\prec=\sort{\mathbf{a}}$.
The discrete uniform distribution is $U\{a, b\}$, whereas the continuous one is $U(a, b)$.
We work in a \emph{static} setting, that is, a fixed point in time $t$.
Hence, throughout the paper, we omit any reference to time without any ambiguity, also to ease the notation.

\paragraph{Unique instruments}
We consider financial instruments of any type, such as fundamental assets (e.g., stocks), derivatives (e.g., swaps, options), securities (e.g., government and corporate bonds), and loans (e.g., mortgages).
We identify a financial instrument having a unitary nominal amount with a set of $m$ (categorical) parameters, namely $\{\phi_1, \ldots, \phi_m\}$.
The parameters $\phi_i$ can vary in their corresponding domains $\domain^{\phi_i}$.
We term the sequence given by concatenating the values of the latter parameters \emph{unique instrument} (UI).
For instance, consider a vanilla option.
A possible parameterization is given by four parameters, i.e., the ticker of the underlying, the payoff type, the strike, and the time-to-maturity.
In this case, a UI is a sequence obtained by concatenating the four values of the previous parameters, each belonging to the associated domain.
Additional parameters can be considered according to the application need (e.g., currency, dividends, counterparty).

\paragraph{Risk factors and risk scenarios}
The UI is associated with a set of random variables $\mathcal{R}=\{R_1, \ldots, R_n\}$, called \emph{risk factors}, which fully determine its price.
For instance, in the case of an option, key risk factors are the price of the underlying $S$, the implied volatility $\sigma$, the interest rate $R$, and the dividend rate $D$.
Each risk factor can be associated with a risk distribution, quantifying the likelihood of different values in the risk factor.
Hence, estimating the risk distribution enables the simulation of risk scenarios for the risk factor.
We denote the dataset of risk scenarios by $\mathcal{F}$.

\paragraph{Unique instruments features}
Features can characterize unique instruments, and we generally refer to them as $\varphi^\ui$.
They can be either observed on the market, thus part of the market data $\mathcal{M}$, or computed by using risk factors $\mathcal{R}$ and risk scenarios $\mathcal{F}$.
Below we provide those relevant to our empirical assessment in \Cref{sec:empirical_assessment}.

\spara{Value.}
Consider a UI, namely \ui.
We indicate its \emph{value} by $\val^\ui$.
The value can be either \emph{marked to market}, in which case the value is quoted on the market, or \emph{marked to model}, when the value is computed using a specific pricing model involving the relevant risk factors.
Pricing models can be chosen according to the institution's internal pricing policy. 
Additionally, they may be calibrated to their reference plain vanilla instruments, and may be based on different numerical approaches (e.g., analytical formulas or Monte Carlo simulations).

\spara{Profit and loss.}
The proposed framework is based on historical scenarios of risk factors $\mathcal{F}$ and the corresponding \emph{profit and loss} (\pl) distribution.
These scenarios simulate past market variations at a fixed point in time $t$ to assess changes in the UI value.
Consider $s$ simulated scenarios. 
We obtain $s$ historical \pl $\historicalpl^\ui_i$, with $i \in [s]$, as follows.
Denote by $\val^\ui_0$ and $\val^\ui_i$ the value of the UI at $t$ and at the $i$-th simulated scenario, respectively.
Hence, we have
\begin{equation}\label{eq:plUI}
    \historicalpl^\ui_i = \val^\ui_i - \val^\ui_0, \quad i \in [s]\,.
\end{equation}
We arrange the previous historical \pl in the vector $\historicalpls^\ui=[\historicalpl^\ui_1, \ldots, \historicalpl^\ui_s]^\top$\,.

\spara{Greeks.}
UIs can be characterized by the sensitivity of their value $\val^\ui$ to changes in the underlying risk factors $\mathcal{R}$.
These sensitivities are known as \emph{Greeks}.
We introduce below the most important Greeks typically used for risk management and trading purposes.
\emph{Delta} is the sensitivity of  $\val^\ui$ with respect to changes in the underlying price $S^\ui$,
\begin{equation}\label{eq:deltaUI}
    \Delta^\ui = \pdv{\val^\ui}{S^\ui}\,.
\end{equation}
\emph{Vega} is the sensitivity of  $\val^\ui$ with respect to changes in the implied volatility of the underlying asset $\sigma^\ui$,
\begin{equation}\label{eq:vegaUI}
    \Vega^\ui = \pdv{\val^\ui}{\sigma^\ui}\,.
\end{equation}
\emph{Gamma} is the sensitivity of Delta with respect to changes in the underlying price $S^\ui$,
\begin{equation}\label{eq:gammaUI}
    \Gamma^\ui = \pdv[2]{\val^\ui}{{S^\ui}}\,.
\end{equation}

\spara{Trading cost.}
Trading financial instruments incurs trading costs. 
It is market practice to measure such costs in terms of sensitivity, i.e., Delta for linear instruments and additionally Vega for options.
Given a UI, namely \ui, we denote such contributions deriving from the buying of a unitary notional amount of \ui by $c^{\Delta^\ui}$ and $c^{\Vega^\ui} \in \reall^+$, respectively.
Hence, the cost for buying a unitary notional amount of \ui is 
\begin{equation}\label{eq:tradingUI}
    c^\ui = c^{\Delta^\ui} \!\!+ c^{\Vega^\ui}\,.
\end{equation}

The computation of $c^{\Delta^\ui}$ and $c^{\Vega^\ui}$ is specific to the nature of \ui.
Henceforth, we provide the cases of stocks, futures on equity indexes, and European call and put options, which are instrumental for the empirical assessment in \Cref{sec:empirical_assessment}.

Denote by \emph{(i)} $\delta$ the market bid/ask spread for the underlying corresponding to \ui, \emph{(ii)} $\delta^q$ the market bid/ask spread for futures, and \emph{(iii)} $\delta^{K}$ the market bid/ask spread for the European options with strike $K$ (expressed as Delta-percentage).
The cost contributions for the corresponding UIs are
\begin{equation}\label{eq:cost_delta_UI}
    c^{\Delta^{\ui}} = \begin{cases}
        \frac{1}{2} \Delta^\ui \delta & \text{if \ui is either a stock, or a call, or a put}, \\
        \frac{1}{2} \delta^{q} & \text{if \ui is a futures};
    \end{cases} 
\end{equation}
and
\begin{equation}\label{eq:cost_vega_UI}
    c^{\Vega^\ui} = \begin{cases}
        \frac{1}{2} \Vega^\ui \delta^{K} & \text{if \ui is either a call or a put}, \\
        0 & \text{otherwise}.
    \end{cases} 
\end{equation}

\paragraph{Portfolio parameterization}
A portfolio is canonically viewed as a collection of financial instruments with nominal amounts different from zero.
The value of a financial instrument may be specified in terms of a unitary value in a given currency times the notional amount.
In our setting, the unitary value is the value $\val^\ui$ of the corresponding UI.
Accordingly, we represent the portfolio as a pair \generalportfolio, where $\asstG=\{\ui_1, \ldots, \ui_n\}$ is the set of UIs in the portfolio, and $\notionalG=[g_1, \ldots, g_n]^\top$ is the integer vector of their corresponding notional amounts, representing long and short positions.
Such portfolio parameterization is particularly convenient during optimization.
First, it enables the search on UIs and nominal amounts to be separated.
Second, it allows the portfolio features and risk measures to be expressed in terms of those of the constituting UIs.
Accordingly, we can compute relevant features and risk measures at the UI level during the initialization phase.
Then, instead of recomputing from scratch the portfolio features and risk measures by means of pricing functions at each iteration, we exploit the stored values during the optimization process thus reducing the computational burden (see \Cref{sec:RATS}).

\paragraph{Portfolio features}
As in the case of the UI, we can associate features with a financial portfolio.
By exploiting the parameterization \generalportfolio, the portfolio feature $\varphi(\asstG, \notionalG)$ can be written as a linear combination of the UIs' feature $ \varphi^{\ui_j}$, where the coefficients $a(g_j)$ depend on the notional amount associated with each UI. 
Mathematically,
\begin{equation}\label{eq:general_ptf_feature}
    \varphi(\asstG, \notionalG) = \sum_j^n a(g_j) \varphi^{\ui_j}\,.
\end{equation}
Starting from \Cref{eq:plUI,eq:deltaUI,eq:vegaUI,eq:gammaUI,eq:tradingUI}, exploiting \Cref{eq:general_ptf_feature}, the portfolio features relevant to our empirical assessment read as
\begin{equation}\label{eq:Pfeat}
    \begin{aligned}
        \val^\asstG &= \sum_j^n g_j \val^{\ui_j}, \quad \text{(Value)}\\
        \historicalpl^\asstG_i &= \sum_j^n g_j \historicalpl^{\ui_j}_i, \quad \text{(\pl for the $i$-th risk scenario, $i \in [s]$)}\\
        \Delta^\asstG &= \sum_j^n g_j \Delta^{\ui_j}, \quad \text{(Delta sensitivity)}\\
        \Vega^\asstG &= \sum_j^n g_j \Vega^{\ui_j}, \quad \text{(Vega sensitivity)}\\
        \Gamma^\asstG &= \sum_j^n g_j \Gamma^{\ui_j}, \quad \text{(Gamma sensitivity)}\\
        c^\asstG &= \sum_j^n c^{\ui_j} \abs{g_j}\,. \quad \text{(Trading cost)}
    \end{aligned}
\end{equation}

As per the UI, we arrange the historical \pl of the portfolio in a vector, namely $\historicalpls^\asstG=[\historicalpl^\asstG_1, \ldots, \historicalpl^\asstG_s]$. 
Additionally, the trading cost involves the absolute notional amount since the trading fee is paid for both long and short trades.

\paragraph{Risk measures}
Starting from the vector of portfolio \pl, we estimate the portfolio \emph{risk measures}, generally denoted by $\rho(\asstG, \notionalG)$.
Below we report those relevant to our application in \Cref{sec:empirical_assessment}.
We remark that our framework is general, and is not limited to using such risk measures.

First, we have the \emph{sample profit and loss} 
\begin{equation}\label{eq:expectedplP}
    \expectedpl^\asstG = \frac{1}{s} \sum_i^s \historicalpl^\asstG_i\,, 
\end{equation}
namely the empirical mean of the vector of portfolio \pl.

Second, the \emph{sample Value at Risk at the $\beta$ percentile}, namely $\varr^\asstG$, obtained using decreasing scenarios weight.
Consider $ \lambda \in (0, 1)$ the decay factor used to give more weight to the most recent scenarios, and define $\alpha = 1 - \beta (1-\lambda^s)$.
Then, setting $i^\star = \lceil \ln{\alpha}/\ln{\lambda} \rceil$, we have
\begin{equation}\label{eq:varP}
    \varr^\asstG = \left[\sort{\historicalpls^\asstG}\right]_{i^\star}\,,
\end{equation}
viz., $\varr^\asstG$
is the $i^\star$-th entry of the vector of sorted portfolio \pl.

\section{Risk-aware trading portfolio optimization}\label{sec:ratpo}
At a high level our goal can be stated as follows.
Consider an initial portfolio \initialportfolio at time $t$ and an objective function $f$, which will be discussed below.
Additionally, consider a set of risk- and trading-based constraints associated with portfolio features and risk measures. 
We want to find the portfolio \totalportfolio optimizing $f$, which is obtained by adding to \initialportfolio a suitable \emph{eligible optimization strategy} (EOS), denoted by \eligibleoptimizationstrategy, complying with the specified constraints.
Mathematically,
\begin{equation}\label{eq:totalptf}
    \totalportfolio \coloneqq \left\langle  \asstP \uplus \asstEOS, \left[\notionalP^\top, \notionalEOS^\top\right]^\top \right\rangle \,.
\end{equation}
We call this task \emph{risk-aware trading portfolio optimization} (RATPO).
In the sequel, we formulate the general RATPO problem.

Looking at RATPO, the first key ingredient is the initial portfolio \initialportfolio.
In our work, \asstP may include any UI, as assumed throughout \Cref{sec:preliminaries}.
The parameterization of such UIs is static and does not depend on the specification of RATPO.
Accordingly, we call $\asstP$ the set of \emph{unique static instruments}.
Analogously, the vector of notional amounts $\mathbf{p} \in \integer^\abs{\asstP}$ is fixed during the optimization process.

The second key ingredient is the objective function $f$.
Consider \emph{(i)} a set of $p$ portfolio risk measures, $\rho_i(\asstTotal, \notionalTotal)$, $i \in [p]$; and \emph{(ii)} a set of $q$ portfolio features $\varphi_j(\asstTotal, \notionalTotal)$, $j \in [q]$.
The family of objective functions considered in RATPO is a general combination of (possibly nonconvex) risk measures and portfolio features. 
Specifically,
\begin{equation}\label{eq:generalf}
    f(\asstTotal, \notionalTotal) \stackrel{(a)}{=} f(\asstEOS, \notionalEOS; \asstP, \notionalP) \coloneqq f\left(\rho_1, \ldots, \rho_p, \varphi_1, \ldots, \varphi_q \right);
\end{equation}
where in $(a)$ we recognize that \initialportfolio is fixed during the optimization.
Additionally, in \Cref{eq:generalf} we omit the arguments for risk measures and the portfolio features to enhance readability.
The specific risk measures, portfolio features, and functional form of $f$ can be chosen according to the application's need, as demonstrated in the empirical assessment in \Cref{sec:empirical_assessment}.

The third key ingredient is a set of $z$ inequality and equality constraints, $\psi_i(\asstEOS, \notionalEOS; \asstP, \notionalP)$, $i \in [z]$.
Similarly to the objective function, each constraint can be a combination of (possibly nonconvex) risk measures and portfolio features, depending on the application need.
The constraints entail the set of \emph{feasible} EOS, denoted by \EOSset, which can be nonconvex.

Next, the fourth key ingredient is the total portfolio \totalportfolio.
Starting from \Cref{eq:totalptf}, the considered \emph{universe} of UIs in RATPO, namely $\uiuniverse$ such that $\asstTotal \subseteq \uiuniverse$,  consists of the multiset union of two distinct sets.
The first, $\asstP$, is the set of UIs in the initial portfolio.
The second, $\uiuniverse^\asstEOS$ such that $\asstEOS \subseteq \uiuniverse^\asstEOS$, is the set of all possible UIs constituting the EOS $\eligibleoptimizationstrategy \in \EOSset$.
The set $\uiuniverse^\asstEOS$ depends on the specification of RATPO since it is entailed by the Cartesian product of the domains $\domain^{\phi_i}$ associated with the parameters $\phi_i$ specifying each UI $\uei_{j}$ that can constitute \asstEOS, where $j \in \left[\abs{\uiuniverse^\asstEOS}\right]$. 
We term the $H_j$ \emph{unique eligible instrument} (UEI), and $\uiuniverse^\asstEOS$ the set of UEIs.
Note that some UIs in \asstP can potentially be in $\uiuniverse^\asstEOS$, depending on the application's need.
This is the reason why we use the multiset union.
Now, we can pose the formal definitions for the UEI and for the EOS.
\bigskip
\begin{definition}[Unique eligible instrument, UEI]\label{def:uei}
    A unique eligible instrument $H_j \in \uiuniverse^\asstEOS$ refers to a financial instrument tradable on the market at the optimization time $t$.
    It is represented as the sequence resulting from the concatenation of a combination in the Cartesian product $\domain^{\phi_1}\times\ldots\times\domain^{\phi_m}$, where $\phi_1,\ldots,\phi_m$ are the parameters determining the UEI, and 
    $\domain^{\phi_1},\ldots,\domain^{\phi_m}$ the corresponding domains.
\end{definition}
\bigskip
\begin{definition}[Eligible optimization strategy, EOS]\label{def:eos}
An eligible optimization strategy \eligibleoptimizationstrategy is a tradable financial strategy at time $t$, complying with a set of specified constraints.
It is a set $\mathcal{H} \subseteq \uiuniverse^\asstEOS$ of UEIs, weighted by the corresponding notional amounts $h_{j} \in \mathbf{h}$, such that $\eligibleoptimizationstrategy \in \EOSset$. 
\end{definition}
\bigskip
At this point, we are ready to formally state the general RATPO problem.
\bigskip
\begin{problem}[Risk-aware trading portfolio optimization, RATPO]\label{prob:RATPO}
Given an initial portfolio \initialportfolio, the total portfolio that optimizes the objective function $f$ is given by $\optimaltotalportfolio \coloneqq \left\langle  \mathcal{P} \uplus \mathcal{H}^\star, \left[\mathbf{p}^\top, \mathbf{h}^{{\star}^\top}\right]^\top \right\rangle $, where   
\begin{equation}\label{eq:generalP}
    \langle \mathcal{H}^\star, \mathbf{h}^\star \rangle = \argmin_{\langle \mathcal{H}, \mathbf{h} \rangle \in \mathcal{E}} \; f(\mathcal{H}, \mathbf{h};  \mathcal{P}, \mathbf{p})\,.
    \tag{P1}
\end{equation}
\end{problem}

The optimization variable in \eqref{eq:generalP}, i.e., the EOS \eligibleoptimizationstrategy, can be mapped into an integer vector as further discussed in \Cref{sec:RATS}.
Thus, Problem \eqref{eq:generalP} can be viewed as an integer optimization problem with possibly nonconvex objective function and constraint set.
We emphasize that nonconvexity can easily originate from the risk measures and portfolio features considered by risk managers and traders in their daily activities.
Hence, the difficulty in solving \eqref{eq:generalP} boils down to the mathematical properties of the objective function in \Cref{eq:generalf} and of the feasible solution set \EOSset.
Establishing a general method that guarantees global convergence for the general RATPO problem is not feasible, and each specification of \eqref{eq:generalP} should be analyzed separately.
The next section exposes a general-purpose algorithm based on particle swarm optimization. 
This method empirically proves to be effective and efficient in solving an instance of the RATPO problem involving the main risk measures and portfolio features used in daily risk management and trading activities (see \Cref{sec:empirical_assessment}).

\section{Risk-aware trading swarm algorithm}\label{sec:RATS}
Meta-heuristics represent general-purpose approaches for solving optimization problems.
Among their features, they do not make restrictive assumptions on the mathematical properties of the objective function and constraints.
This approach allows risk managers and traders to specify arbitrarily complex objective functions of the form in \Cref{eq:generalf} arising from the need to satisfy trading, risk management, and regulatory objectives.
In addition, this aspect allows specifying complex constraints set, and thus guiding the optimizer's search toward solutions to problem \eqref{eq:generalP} having precise financial meaning.  
In addition, they do not leverage the gradient during the optimization.
Therefore, they are amenable to handling optimization variables of different natures.
Finally, they efficiently explore the solution space by balancing the exploration-exploitation trade-off (see. \cite{Erwin2023meta}).

The compromise for this flexibility is the lack of convergence guarantees to a global optimum within a finite time. 
These algorithms may become trapped in local optima, making it challenging to assess their proximity to the global optimum. 
Nonetheless, for risk managers and traders---who typically possess deep domain knowledge that can inform the optimization problem through constraints---a consistent reduction in the objective function remains valuable.

In our work, we build upon \emph{particle swarm optimization} (PSO), originally introduced by \cite{KE95}.
PSO is inspired by the flocking behavior of birds and fishes, and represents one of the most popular meta-heuristics based on the \emph{swarm intelligence paradigm} by \cite{kennedy2006swarm}.
In PSO, each particle in the swarm explores positions, i.e., candidate solutions of the optimization problem, in an $n$-dimensional space.
Specifically, denoting by $n^p$ the number of particles, the particle is described by the position and velocity vectors, $\x_i$ and $\myv_i$, $i \in [n^p]$, respectively.
The algorithm proceeds by iteratively updating the particles' position and velocity vectors.
We denote by ${\pbest}^{k}_i$ and ${\gbest}^k$ respectively \emph{(i)} the best position visited by the $i$-th particle at iteration $k$, and \emph{(ii)} the best global position visited by the swarm at iteration $k$, respectively.
Given \emph{(i)} the \emph{inertia weight} $w \in \reall^+$, \emph{(ii)} the \emph{personal} and \emph{social coefficients} \cpers\ and \csoc\ in $\reall^+$ respectively, and \emph{(iii)} the vectors \rone and \rtwo drawn from $U(0,1)$, the particle update recursion is
\begin{equation}\label{eq:updatePSOp}
    \begin{cases}
        \myv_i^{k+1} &= w \myv_i^k + \underbrace{\cpers \rone^{k+1} \circ ({\pbest}^{k}_i - \x_i^k)}_{\text{personal component}} + \underbrace{\csoc \rtwo^{k+1} \circ ({\gbest}^{k} - \x_i^k)}_{\text{social component}}\,, \\  
        \x^{k+1}_i &= \x^k_i + \myv^{k+1}_i\,.
    \end{cases}
    \tag{R1}
\end{equation}
By building on the key concepts of UEI and EOS, we develop a special version of PSO that leverages the parameterization \eligibleoptimizationstrategy and enables parallel computation over the particles to efficiently solve the RATPO problem \eqref{eq:generalP}.
Our method, termed \emph{risk-aware trading swarm} (\algorithmname), is described below.

\spara{Initialization.}
As a first step, \algorithmname runs an initialization phase.
Consider that the cardinality of \UEIsuniverse is $n$, and that the names of the UEIs are sorted in lexicographic order.
Since each $\uei_j \in \UEIsuniverse$ corresponds to a unique combination of the representing parameters' values, we can simply identify the $\uei_j$ with its index $j \in [n]$\footnote{The same mapping can be applied to the notional amounts to handle the case in which the notional amount domain is undersampled, that is, the range of the corresponding integers is discretized. We use such a mapping in our empirical assessment in \Cref{sec:empirical_assessment}.}.
Hence, the particle position can be represented as a $2m$-dimensional integer vector
\begin{equation}\label{eq:positionRATS}
    \x \coloneqq \left[\underbrace{x_1, \ldots, x_m}_{\text{UEI indices}}, \underbrace{x_{m+1}, \ldots, x_{2m}}_{\text{notional amounts}}\right]^\top\!, \quad \text{with } m\leq n.
\end{equation}
The particle position in \Cref{eq:positionRATS} entails an optimization strategy \eligibleoptimizationstrategy.
If $\eligibleoptimizationstrategy \in \EOSset$, meaning the optimization strategy is \emph{eligible}, the particle position is called \emph{feasible}.
Conversely, if $\eligibleoptimizationstrategy \notin \EOSset$, the particle position is called \emph{unfeasible}.
Furthermore, the number of distinct UEIs represented in \x, i.e., $\abs{\mathcal{H}}$, is at most $m$.
This is because, in principle, the first $m$ entries of \x may include repeated indices.
Consequently, \algorithmname enables users to easily set an upper bound of $m$ on the number of financial instruments that can be selected, a helpful feature from an application viewpoint.

\algorithmname initializes the entries $x_j$, $j \in [m]$, by sampling from $U\{b_j, u_j\}$, where $b_j$ and $u_j \in \nat$, such that $1 \leq b_j < u_j \leq n$. 
Hence, users can easily specify the relevant UEIs for each entry $x_j$ through $b_j$ and $u_j$.
Next, \algorithmname initializes the entries $x_{j+m}$, $j \in [m]$, by sampling from $U\{\ell_j, t_j\}$, where $\ell_j$ and $t_j$ are minimum and maximum notional amounts, respectively.
Thus, $\ell_j$ and $t_j$ act on the UEIs corresponding to the indices from $b_j$ to $u_j$ within $\UEIsuniverse$.
As far as the velocity vector is concerned, \algorithmname initializes its entries by sampling from $U(v^{\mathrm{min}}, v^{\mathrm{max}})$, where $v^{\mathrm{min}}$ and $v^{\mathrm{max}}$ in $\reall$ are two hyper-parameters corresponding to the minimum and maximum velocity.

Additionally, starting from the input datasets of risk factors $\mathcal{R}$, market data $\mathcal{M}$, and risk scenarios $\mathcal{F}$, \algorithmname precomputes the $p$ features of the UEIs relevant to the optimization problem, namely $\varphi_q^{\uei_j}$ with $q \in [p]$ and $j \in [n]$.
This is possible since the features of the UEIs do not change during the optimization process.
This way, \algorithmname lowers the computational burden of the optimization process.
Indeed, at each iteration $k$, to assess the goodness and feasibility of the particles' positions $\x_i^k$, $i \in [n^p]$, we need to compute the features of the optimization strategy entailed by $\x_i^k$ since they are necessary for the evaluation of \Cref{eq:generalf} and of the constraints $\psi_\ell(\asstEOS,\notionalEOS;\asstP, \notionalP)$, $\ell \in [z]$ (cf. \Cref{sec:preliminaries}).
In general, assessing a feature of an optimization strategy involves evaluating a (nonlinear) function, which can be computationally expensive.
On the contrary, \algorithmname efficiently exploits the optimization strategy parameterization which allows us to use \Cref{eq:general_ptf_feature}.
Hence, at each iteration and for each particle, the computational cost of the optimization strategy feature evaluation is linear, as it requires $2m-1$ operations.

Next, \algorithmname evaluates the \emph{fitness} of the initial positions of the particles to set $\pbest_i$ and $\gbest$.
Starting from \Cref{eq:generalf}, we define the fitness function as
\begin{equation}\label{eq:fitnessf}
    \bar{f}(\asstEOS,\notionalEOS;\asstP, \notionalP) \coloneqq f(\asstEOS,\notionalEOS;\asstP, \notionalP) + \sum_\ell^z \lambda_\ell \left[\psi_\ell(\asstEOS,\notionalEOS;\asstP, \notionalP)\right]_+\,.
\end{equation}
The second term in \Cref{eq:fitnessf} represents the weighted sum of the violations of the inequality and equality constraints, where $\lambda_j \in \reall^+$ are penalty hyper-parameters.
If the position is feasible, the value of the fitness function is equal to that of the objective function.
Conversely, it is penalized by the weighted sum of constraints' violations. 
Henceforth, we refer to the fitness function evaluated at the optimization strategy entailed by a particle position \x as $\bar{f}\at[\big]{\x}$.

\spara{Recursion.}
After the initialization phase, \algorithmname runs the recursion in \eqref{eq:updatePSOp} to update the position and the velocity of the particle.
In our setting, the particle position $\x_i^k$ takes on values in $\integer^{2m}$, whereas the velocity vector $\myv_i^{k+1} \in \reall^{2m}$.
Hence, after the second update in \eqref{eq:updatePSOp}, we apply rounding to ensure that $\x_i^{k+1}$ is a vector of integers and its entries are within the specified ranges determined by $\{b_j,\, u_j,\, \ell_j,\, t_j\}$, $j \in [m]$.
At this point, \algorithmname evaluates $\bar{f}\at[\big]{\x_i^{k+1}}$, and accordingly determines ${\pbest}_i^{k+1}$.
\algorithmname runs the above steps in parallel over the particles, thus reducing the computational time, and allowing the user to choose larger values of $n^p$ for better exploration.
Next, \algorithmname updates the global best position if it finds a new position that improves the fitness function evaluated at the global best position more than $\tau^f$, where $\tau^f \in \reall^+$ is a small significance threshold.
Specifically, the condition reads as
\begin{equation}\label{eq:updatecond}
    \bar{f}\at[\Big]{{\gbest}^{k}} - \min_{i\in [n^p]} \bar{f}\at[\Big]{{\pbest}_i^{k+1}}  > \tau^f\,.   
\end{equation}
In case \Cref{eq:updatecond} is not satisfied, \algorithmname increases a counter $k^{\mathrm{stall}} \in \nat$ which keeps track of the number of stall iterations.
Finally, following \cite{Shi98}, we dynamically adjust the inertia weight, according to a linearly decaying rule. 
Given minimum and maximum values, $w^{\min}$ and $w^{\max}$, respectively; and setting $w^1 = w^{\max}$, the rule reads as
\begin{equation}\label{eq:inertia_update}
    w^{k+1} = w^{\max} - \frac{k}{k^{\max}} \left(w^{\max}- w^{\min}\right)\, ,
\end{equation}
where $k^{\max}$ is the maximum number of iterations.

\spara{Stopping criteria.}
\algorithmname exits the recursion either \emph{(i)} if the maximum number of iterations $k^{\max}$ is reached, 
or \emph{(ii)} if the global best position is not updated for a specified number of consecutive stall iterations $k^\mathrm{stall} \geq k^\mathrm{max \, stall}$, 
or \emph{(iii)} if the fraction $\chi$ of particles exhibiting a personal best position equal to the global best one exceeds a concentration threshold $\tau^p \in (0,1)$.

\medskip
Algorithm~\ref{algo:RATS} summarizes the overall procedure.
We emphasize that \algorithmname can be easily equipped by the user with different existing variants of PSO.

\begin{algorithm}[H]\label{algo:RATS}
\caption{Risk-aware trading swarm (\algorithmname)}
\DontPrintSemicolon 
\SetKwBlock{DoParallel}{do in parallel}{end}
\KwIn{risk factors $\mathcal{R}$, market data $\mathcal{M}$, risk scenarios $\mathcal{F}$, initial portfolio $\initialportfolio$, unique eligible instruments $\{H_j\}$ with $j \in [n]$, number of particles $n^p$, minimum/maximum UEI indices/notional amounts $\{b_j, \, u_j, \ell_j, \, t_j\}$ with $j \in [m]$, minimum velocity $v^{\min}$, maximum velocity $v^{\max}$, minimum inertia weight $w^{\min}$, maximum inertia weight $w^{\max}$, small significance threshold $\tau^f$, concentration threshold $\tau^p$,  maximum number of iterations $k^{\max}$, number of consecutive stall iterations $k^{\mathrm{max\, stall}}$}
\KwOut{\gbest}
$w \gets w^{\max}$\;
$k^{\mathrm{stall}} \gets 0$\;

\For{$j \leftarrow 1$ \KwTo $n$}{
    \For{$q \leftarrow 1$ \KwTo $p$}{
        $\varphi^{\uei_j}_q \gets$ Compute it by using $\mathcal{R},\, \mathcal{M},\, \mathcal{F}$\;
    }
}

\For{$i \leftarrow 1$ \KwTo $n^p$}{
    $\x_i^0 \gets $ Sample the first $m$ entries from $U\{b_j, u_j\}$, the rest from $U\{\ell_j, t_j\}$, with $j \in [m]$\;
    $\myv_i^0 \gets $ Sample the entries from $U\left(v^{\min},\, v^{\max}\right)$\;
    $\pbest_i^0 \gets \x_i^0$\; 
    $\bar{f}\at[\big]{\x_i^0} \gets $ Apply \Cref{eq:fitnessf}\;
    $\bar{f}\at[\big]{\pbest_i^0} \gets \bar{f}\at[\big]{\x_i^0}$\; 
}
${\gbest}^0 \gets \argmin_{\pbest_i^0} \bar{f}\at[\Big]{\pbest_i^0}$\;
$\bar{f}\at[\big]{{\gbest}^0} \gets $ Apply \Cref{eq:fitnessf}\;
\While{$k<k^{\max} \; \textbf{and} \; k^{\mathrm{stall}}<k^{\mathrm{max \, stall}} \; \textbf{and} \; \chi < \tau^p $}{
    ${\gbest}^k \gets {\gbest}^{k-1}$\;
    $\mathbf{r}_1,\,\mathbf{r}_2 \gets $ Sample the entries from $U\!\left(0,\, 1\right)$\;
    \DoParallel{
    $\pbest_i^k \gets \pbest_i^{k-1}$\;
    $\myv_i^k, \x_i^k \gets $ Apply \eqref{eq:updatePSOp}\;
    $\bar{f}\at[\big]{\x_i^k} \gets $ Apply \Cref{eq:fitnessf}\;
    \If{$\bar{f}\at[\big]{\x_i^k} < \bar{f}\at[\big]{\pbest_i^k}$}{$\pbest_i^k \gets \x_i^k$\;}
    }
    \eIf{\Cref{eq:updatecond}}{
    ${\gbest}^k \gets \argmin_{\pbest_i^k} \bar{f}\at[\Big]{\pbest_i^k}$\;
    $k^{\mathrm{stall}} \gets 0$\;
    }{$k^{\mathrm{stall}} \gets k^{\mathrm{stall}} +1 $\;}
    $w^{k+1} \gets $ Apply \Cref{eq:inertia_update}\;
}
\end{algorithm}

\section{Empirical assessment}\label{sec:empirical_assessment}
This section provides the empirical assessment of the proposed approach via two different applications.
The first, given in \Cref{subsec:toy_portfolio}, concerns a simplified setting that is instrumental to show that \algorithmname described in \Cref{sec:RATS} can find a solution in the optimal solution set corresponding to the lowest objective function value.
Indeed, although the initial portfolio is made of hundreds of instruments of different kinds, the cardinality of the solution space is tractable (roughly of the order of $10^8$), and it is possible to find the optimal objective function value and the associated optimal solution set via brute force.
This first application is also helpful to point out the main differences between the proposed approach and a Markowitz-like approach.

The second application concerns a more real setting where the initial portfolio is still the same, but the set of instruments for building the EOS is larger, and the cardinality of the solution space is roughly of the order of $10^{110}$.
In this case, finding the optimal value of the objective function via brute force is not computationally feasible.

\subsection{Experimental setting}\label{subsec:expset}
Throughout the section, we use the diacritics $\widetilde{\;}$ (tilde) when we refer to the quantities involved in the RATPO problem to emphasize that we are in a specific application setting.
Below are the specifications for the initial portfolio, for the objective function $\widetilde{f}$, and for the constraints determining the feasible set of EOS, namely $\widetilde{\EOSset}$, used in both applications.
These specifications are needed to pose the RATPO problem \eqref{prob:optprob_cs} tackled in our empirical assessment, that is, a particular case of \eqref{eq:generalP}.

\paragraph{Initial portfolio}
We consider a real trading portfolio \initialportfolio including $127$ financial instruments in $6$ different currencies on $87$ underlying equity stocks and indices, as reported in \Cref{tab:intial_ptf}. 
The initial portfolio composition, features, and risk measures reflect the trader's view at $t$, i.e., as of September $28$, $2018$.
The initial portfolio static UIs are determined by parameters $\phi_i$, such as underlying, payoff type, maturity, and strike (see \Cref{sec:preliminaries}).
We consider five features for the static UIs, namely the value, the \pl, and the sensitivities Delta, Vega, and Gamma.
The value at $t$ is marked to market. 
Conversely, in the case of risk scenarios, for stocks and futures, the value is marked to market, whereas for options, it is marked to model, i.e.,
computed according to specific pricing models.\footnote{We use the model by \cite{black1973pricing} for European options, and the approximation provided by \cite{barone1987efficient} for American options. We remark that our approach only requires the usage of non-sophisticated pricing models to manage plain vanilla UEIs.
Then, the other features are computed by using \Cref{eq:plUI,eq:deltaUI,eq:vegaUI,eq:gammaUI}.
More precisely, the sensitivities Delta and Gamma are based on a $1\%$ multiplicative shock of the underlying asset. 
The sensitivity Vega is based on a $1\%$ additive shock of the implicit volatility of the underlying. 
All the features are either observed or computed in their reference currencies and then converted to EUR currency.
The number of considered risk scenarios is $s=250$.}

Starting from the features of the static UIs and exploiting \Cref{eq:Pfeat}, we compute the initial portfolio value $\val^\asstP$, its \pl in the risk scenarios, and the sensitivities $\Delta^\asstP$, $\Vega^\asstP$, $\Gamma^\asstP$.
\Cref{tab:intial_ptf} provides $\val^\asstP$, $\Delta^\asstP$, $\Vega^\asstP$, and $\Gamma^\asstP$.
Additionally, the \pl distribution is given in \Cref{fig:cs1_init_vs_best_total}.

\begin{table}[t]
    \centering
    \caption{Composition, values and sensitivities of \initialportfolio, and its constituents.
    Specifically, we report aggregate values at the type level for the latter.
    Market data as of September $2018$, the $28^{\mathrm{th}}$.}
    \sisetup{
        detect-mode,
        table-format		    = 8.2
        }    
    \begin{tabular}{lcSSSS}
    \toprule
        Instrument & {Number} & \text{$v^{\mathcal{P}}$ (EUR)} & \text{$\Delta^{\mathcal{P}}$ (EUR)} & \text{$\mathcal{V}^{\mathcal{P}}$ (EUR)} & \text{$\Gamma^{\mathcal{P}}$ (EUR)} \\
        \midrule
        Stocks              & 75 & 13465008  & 134650     & 0      & 0      \\
        Futures             & 14 & 0         & -232792    & 0      & 0      \\
        European options    & 10 & 99177     & -20042     & 12159  & 7520   \\ 
        American options    & 28 & 236664    & 73529      & 67391  & 7424   \\
        \midrule
        Total               & 127& 13800849  & -44655     & 79550  & 14944  \\ 
        \bottomrule
    \end{tabular}
    \label{tab:intial_ptf}
\end{table}

\paragraph{Objective function} 
The objective function for the two applications is the same.
It consists of the combination of two portfolio risk measures and one portfolio feature.
Let us indicate with $\mathrm{P\&L}^{\mathcal{P}}$ the profit we would make in one day by investing \initialportfolio at the risk-free rate corresponding to the rebalancing date.
As risk measures of interest for the \totalportfolio given in \Cref{eq:totalptf}, we consider the sample P\&L, denoted by \expectedplT, and the \varrT with $\beta=1\%$, defined in \Cref{eq:expectedplP,eq:varP}, respectively.
In our experiments, we consider the \varrT as negative-valued.
The considered feature is the transaction cost for implementing \totalportfolio, defined in \Cref{eq:Pfeat}.
Please notice that the transaction cost is only determined by the EOS \eligibleoptimizationstrategy since \initialportfolio is fixed.
Accordingly, we denote it by $c^\asstEOS$.
Hence, the objective function reads as

\begin{equation}\label{eq:obj_fun_cs}
    \widetilde{f}(\asstEOS, \notionalEOS; \asstP, \notionalP) = \frac{\expectedplT - \mathrm{P\&L}^{\mathcal{P}} - c^{\asstEOS}}{\varrT - c^{\asstEOS}}\,.
\end{equation}

\Cref{eq:obj_fun_cs} provides a cost-adjusted risk measure of \totalportfolio.
The sign of the cost term can be easily understood by noticing that \emph{(i)} at the numerator the cost term erodes the excess P\&L of \totalportfolio w.r.t. \initialportfolio, and \emph{(ii)} at the denominator it represents an additional loss to that conveyed by the \varrT (which is typically already negative).

Ideally, we would like to maximize the numerator and to vanish the denominator.
Since at the denominator we approach zero from the left, the more negative the value of $\widetilde{f}$, the better \totalportfolio.
Hence, in our empirical assessment, we aim at minimizing the objective function in \Cref{eq:obj_fun_cs}.

We remark that the objective function in \Cref{eq:obj_fun_cs} is a particular case of the general objective in \Cref{eq:generalf}, and that our approach is not limited to it.

\paragraph{Constraints}
We consider Delta, Vega, and Gamma sensitivity constraints.
These kinds of constraints are common in day-to-day risk management and trading activities.
We remark that the proposed approach is general and not limited to this specific choice for the constraints.
The sensitivity constraints are defined according to three portfolio features.
Indicate with $\Delta^{\mathcal{P}}$ the monetary value of the delta sensitivity for \initialportfolio, and the same for the vega and gamma sensitivities, $\mathcal{V}^{\mathcal{P}}$ and $\Gamma^{\mathcal{P}}$, respectively.
Analogously, we indicate by $\Delta^{\mathcal{H}}, \, \mathcal{V}^{\mathcal{H}}$, and $\Gamma^{\mathcal{H}}$ the monetary value of the considered sensitivities for \eligibleoptimizationstrategy.
Hence, given $\tau^{\Delta},\, \tau^{\mathcal{V}},\, \tau^{\Gamma} \in \reall^+$, the sensitivity constraints read as
\begin{equation}\label{eq:gconstraint_cs}
    \begin{aligned}
        \widetilde{\psi}^{\Delta}(\asstEOS, \notionalEOS; \asstP, \notionalP) \coloneqq& \abs{\Delta^{\mathcal{H}}} - \tau^{\Delta} \abs{\Delta^{\mathcal{P}}} \leq 0 \,;\\
        \widetilde{\psi}^{\Vega}(\asstEOS, \notionalEOS; \asstP, \notionalP) \coloneqq& \abs{\Vega^{\mathcal{H}}} - \tau^{\Vega} \abs{\Vega^{\mathcal{P}}} \leq 0\,;\\
        \widetilde{\psi}^{\Gamma}(\asstEOS, \notionalEOS; \asstP, \notionalP) \coloneqq& \abs{\Gamma^{\mathcal{H}}} - \tau^{\Gamma} \abs{\Gamma^{\mathcal{P}}} \leq 0\,.
    \end{aligned}
\end{equation}
Given the linearity of the sensitivities in the vector of notional amounts as per \Cref{eq:Pfeat}, the constraints in \Cref{eq:gconstraint_cs} are linear inequality constraints.

\paragraph{Unique eligible instruments}
We consider a list of $u \in \nat$ underlyings, sorted in lexicographical order.
Each underlying is either a stock or a stock index.
If the underlying is a stock, we consider as UEIs European vanilla options written on the stock and the stock itself.
Conversely, we consider European vanilla options and futures written on the stock index to be UEIs.
To formally implement \Cref{def:uei} in our empirical assessment, we consider the following parameters:
\begin{squishlist}
    \item $\nu \in [u]$ the position of the underlying  within the list of underlyings; 
    \item $\omega \in \{\mathrm{c},\mathrm{p},\mathrm{q},\mathrm{s}\}$  the type of the UEI, i.e., \enquote{$\mathrm{c}$} for call, \enquote{$\mathrm{p}$} for put, \enquote{$\mathrm{q}$} for futures, \enquote{$\mathrm{s}$} for stock;
    \item $K \in \{0.10,\,0.25,\,0.50\}$ the strike expressed as $\Delta$-percentage;
    \item $T \in \domain^{T_1} \cup \ldots \cup \domain^{T_u}$ the time-to-maturity in days, where $\domain^{T_\ell}$ is the domain corresponding to the underlying at the position $\nu=\ell$ within the list of underlyings.
\end{squishlist}
Please refer to \Cref{tab:ueis_features} in Appendix \ref{app:ueis_features_sn} for further details regarding the parameters
associated with the UEIs.
Additionally, recall that we denote by $n$ the cardinality of $\UEIsuniverse$, i.e., the number of considered UEIs.
At this point, we are ready to implement \Cref{def:uei} in our application setting.

\smallskip
\begin{impdefinition}[Unique eligible instrument, UEI]\label{def:UEI_application}
A unique eligible instrument $H_j \in \UEIsuniverse$, $j \in [n]$, is the string obtained by the concatenation of the value of the parameters
\begin{equation}
    \begin{cases}
        \nu_j, \text{and } \omega_j\,, &\quad \text{for stocks;}\\
        \nu_j,\, \omega_j,\,K_j, \text{and } T_j\,, &\quad \text{for European options and futures.}\\
    \end{cases}
\end{equation}
\end{impdefinition}
\smallskip

\paragraph{Eligible optimization strategy}
Next, starting from \Cref{def:UEI_application}, considering the sensitivity constraints in \Cref{eq:gconstraint_cs}, we can specify \Cref{def:eos} in our setting.
\bigskip
\begin{impdefinition}[Eligible optimization strategy]\label{def:eos_set_cs}
    Given $\Delta^{\mathcal{P}}$, $\mathcal{V}^{\mathcal{P}}$, $\Gamma^{\mathcal{P}}$, an eligible optimization strategy \eligibleoptimizationstrategy is a set $\mathcal{H} \subset \UEIsuniverse$ of $m \leq n$ UEIs, weighted by the corresponding notional amounts $\notionalEOS=[h_1, \ldots, h_m]^\top$, satisfying the sensitivity constraints in \Cref{eq:gconstraint_cs}.
\end{impdefinition}

\smallskip
As mentioned in \Cref{sec:RATS}, thanks to our parameterization we can focus on those EOS having a precise structure on both the set of selected UEIs $\asstEOS$ and corresponding notional amounts in $\notionalEOS$.
This structure is meant to reflect the prior knowledge of traders and risk managers.
Recall that $u$ is the number of underlyings and consider the UEIs sorted in lexicographical order.
As an example, concerning the structure of $\asstEOS$, in our applications, we require that the EOS contains at most $n^i$ UEIs for each underlying, thus implying a cardinality upper-bound $\abs{\mathcal{H}}\leq m = n^i \, u\,$ for the strategy (see \Cref{sec:RATS}). 
Specifically, in our experiments, we consider the case where $n^i=3$.
Additionally, if the underlying is a stock we require that $\mathcal{H}$ contains (at most) two vanilla options written on the underlying and the stock itself; if a stock index, (at most) two vanilla options and a futures written on the underlying.
This is useful in showing how it is possible to control the size of the EOS---which is important from the operational viewpoint---and how to favor balanced solutions.

Starting from \Cref{eq:positionRATS}, we know that the particle position \x is a $2m$-dimensional integer vector.
The first $m$ entries correspond to the indices of the UEIs, and the second to their notional amounts. 
Hence, to impose the above structure onto $\asstEOS$, we partition the first $m$ entries of \x into triplets, each corresponding to the $n^i=3$ UEIs for a specific underlying:
\begin{equation}\label{eq:partitionxH}
    \underbrace{x_{1,1},\,x_{1,2},\,x_{1,3}}_{\text{$1^{\mathrm{st}}$ underlying }},\,\ldots,\, \underbrace{x_{\ell,1},\,x_{\ell,2},\,x_{\ell,3}}_{\text{$\ell^{\mathrm{th}}$ underlying}}, \ldots,\,\underbrace{x_{u,1},\,x_{u,2},\,x_{u,3}}_{\text{$u^{\mathrm{th}}$ underlying}}\,.
\end{equation}
Hereinafter, to enhance readability and with a slight abuse of notation, we use a double indexing: the first on triplets, the second on the elements of a triplet.
Looking at \Cref{eq:partitionxH}, we have that that the $\ell$-th triplet corresponds to those UEIs $\uei_j \in \UEIsuniverse$, $j \in [n]$, having $\nu_j = \ell$.
Then, within each triplet, we force the first two entries to be European options and the third to be either a stock or a stock index, depending on the nature of the $\ell$-th underlying.
Again, referring to the $\ell$-th triplet within \Cref{eq:partitionxH}, this means that the $x_{\ell,1}$ and $x_{\ell,2}$ correspond to those $\uei_j$ with $\nu_j=\ell$ and $\omega_j \in \{\mathrm{c}, \mathrm{p}\}$.
Analogously, $x_{\ell,3}$ to those $\uei_j$ with $\nu_j=\ell$ and $\omega_j = \text{\enquote{$\mathrm{s}$}}$ if the $\ell$-th underlying is a stock, $\omega_j = \text{\enquote{$\mathrm{q}$}}$ if a stock index.
Hence, for each entry $x_{\ell,i}$, where $\ell \in [n]$ and $i \in [n^i]$, we specify accordingly the lower and upper bound $[b_{\ell,i}, u_{\ell,i}]$.
If $i \in \{1,\,2\}$, then $b_{\ell,i}$ is the lower bound of the range of indices corresponding to the UEIs with $\nu_j=\ell$ and $\omega_j=\mathrm{c}$, whereas $u_{\ell, i}$ is the upper bound of the range of indices corresponding to the UEIs with $\nu_j=\ell$ and $\omega_j=\mathrm{p}$ (notice that \enquote{$\mathrm{c}$} precedes \enquote{$\mathrm{p}$} and the underlyings are lexicographically sorted).
If $i = 3$ and the $\ell$-th underlying is a stock, then $b_{\ell,i}=u_{\ell,i}$ is the index of the (only) UEI with $\nu_j=\ell$ and $\omega_j=\mathrm{s}$ (see \Cref{def:UEI_application}).
Instead, if the $\ell$-th underlying is a stock index, then $b_{\ell,i}$ is the lower bound of the range of indices corresponding to the UEIs with $\nu_j=\ell$ and $\omega_j=\mathrm{q}$, $u_{\ell,i}$ the upper bound.

Concerning the notional amounts in \notionalEOS, the partition in \Cref{eq:partitionxH} entails a partition on the second $m$ entries of \x corresponding to the notional amounts:
\begin{equation}\label{eq:partitionxh}
    \underbrace{x_{m+1,1},\,x_{m+1,2},\,x_{m+1,3}}_{\text{$1^{\mathrm{st}}$ underlying }},\,\ldots,\, \underbrace{x_{m+\ell,1},\,x_{m+\ell,2},\,x_{m+\ell,3}}_{\text{$\ell^{\mathrm{th}}$ underlying}}, \ldots,\,\underbrace{x_{m+u,1},\,x_{m+u,2},\,x_{m+u,3}}_{\text{$u^{\mathrm{th}}$ underlying}}\,.
\end{equation}
In our empirical assessment, we consider minimum and maximum notional amounts for each entry $x_{m+\ell,i}\,$, where $\ell \in [u]$ and $i \in [n^i]$, denoted by $\ell_{\ell,i}$ and $t_{\ell,i}\,$, respectively.
\Cref{tab:ueis_features} provides the latter values.
Furthermore, the ranges $[\ell_{\ell,i},\, t_{\ell,i}]$ are discretized according to the procedure detailed in Appendix \ref{app:ueis_features_sn}.
This is instrumental in showing how easy it is to handle cases where there is an operational need to trade batches of shares.

Starting from \Cref{def:eos_set_cs}, considering the structure for the set of UEIs and notional amounts detailed above, we denote by $\widetilde{\EOSset}$ the resulting EOS set.

\paragraph{RATPO problem}
Starting from the objective function in \Cref{eq:obj_fun_cs}, considering \Cref{def:eos_set_cs} and $\widetilde{\EOSset}$ given as above, we pose the implementation of the general RATPO \Cref{prob:RATPO} we aim at solving in the applications in \Cref{subsec:toy_portfolio,subsec:real_portfolio}.
\bigskip
\begin{impproblem}
Given an initial portfolio \initialportfolio, the total portfolio that optimizes the objective function in \Cref{eq:obj_fun_cs} subject to the sensitivity constraints in \Cref{eq:gconstraint_cs} is given by $\optimaltotalportfolio \coloneqq \left\langle \mathcal{P} \uplus \mathcal{H}^\star, \left[\mathbf{p}^\top, \mathbf{h}^{\star^\top}\right]^\top \right\rangle $, where   
\begin{equation}\label{prob:optprob_cs}
    \left\langle \mathcal{H}^\star, \mathbf{h}^\star \right\rangle = \argmin_{\langle \mathcal{H}, \mathbf{h} \rangle \in \widetilde{\EOSset}} \; \frac{\expectedplT - \mathrm{P\&L}^{\mathcal{P}} - c^{\asstEOS}}{\varrT - c^{\asstEOS}}\,.
    \tag{$\widetilde{\mathrm{P1}}$}
\end{equation}
We denote the set of optimal solutions by $\widetilde{\mathcal{E}}^\star$.
\end{impproblem}
\smallskip
Unfortunately, \eqref{prob:optprob_cs} is nonconvex.
First, the fractional objective function contains the \varrT at the denominator, which is inherently nonconvex (\cite{Lwin2017mean} and refs. therein).
Second, the EOS set $\widetilde{\EOSset}$ is nonconvex as well due to the structure we impose on \asstEOS (see paragraph \enquote{Eligible optimization strategy} above).
Indeed, from a canonical optimization perspective, imposing this structure boils down to a group-wise cardinality constraint on the $\ell_0$-norm\footnote{The $\ell_0$-norm of a vector is a pseudo-norm defined as the number of nonzero elements in the vector.} of \notionalEOS, which is nonconvex.

Although we do not exclude the possibility of deriving a convex surrogate of \Cref{eq:obj_fun_cs} (\cite{Wozabal2010} and refs. therein), and approximating with differentiable functions the (group-wise) cardinality constraint as proposed by \cite{malek2016successive}, we believe that such an investigation for deriving an alternative (gradient-based) optimization approach is beyond the scope of our empirical assessment, aimed at demonstrating the relevance of the RATPO problem and the general applicability of RATS given in Algorithm \ref{algo:RATS}.

\subsection{Comparison with the Markowitz-like approach}\label{subsec:M_approach}
Before delving into the case studies, we relate our approach to a possible variation of the Markowitz mean-variance optimization framework, focusing on the challenges to be addressed for solving \eqref{prob:optprob_cs} with a Markowitz-like approach.

\paragraph{The M-approach}
In his seminal work, \cite{Mar1952} defines the portfolio selection problem in terms of expected returns and the variance of returns of $n^a$ assets composing an investment universe $\mathcal{U} \coloneqq \{A_1, \ldots, A_{n^a}\}$.
Indicate with $\mathbf{y}=[y_1, \ldots, y_{n^a}]^\top, \, \mathbf{y} \in \reall^{n^a}$, the vector of portfolio weights for the assets in $\mathcal{U}$.
Assume that the portfolio is fully invested, i.e., $\sum_{j \in [n^a]} y_j=1$.
Denote by $\mathbf{z}\coloneqq [z_i, \ldots, z_{n^a}]^\top, \, \mathbf{z}\in \reall^{n^a}$, the vector of assets returns.
Indicate with $\boldsymbol{\mu}=\mathbb{E}[\mathbf{z}], \boldsymbol{\mu} \in \reall^{n^a}$, the assets expected returns, and assume the covariance matrix $\boldsymbol{\Sigma}=\mathbb{E}\left[(\mathbf{z}-\boldsymbol{\mu})(\mathbf{z}-\boldsymbol{\mu})^\top\right], \, \boldsymbol{\Sigma} \in \reall^{{n^a}\times{n^a}}$, to be positive-definite.
Hence, the portfolio expected return and variance read as:
\begin{equation}\label{eq:mu_sigma}
    \mathrm{\emph{(i)}}\quad \mu(\mathbf{y})=\boldsymbol{\mu}^\top \mathbf{y}, \quad \quad \mathrm{\emph{(ii)}} \quad \sigma^2(\mathbf{y})=\mathbf{y}^\top \boldsymbol{\Sigma} \mathbf{y}\,. 
\end{equation}

By indicating with $\gamma \in \reall^+$ the investor risk aversion, exploiting \emph{(i)} and \emph{(ii)} in \Cref{eq:mu_sigma}, the Markowitz portfolio optimization problem reads as the following quadratic programming problem:
\begin{equation}\label{prob:Markowitz}
    \begin{aligned}
        \mathbf{y}^\star =&  \argmin_{\mathbf{y}} \frac{\gamma}{2} \sigma^2(\mathbf{y}) - \mu(\mathbf{y})\,,\\
        &\text{subject to} \quad \sum_{j \in [n^a]} y_j=1\,.
    \end{aligned}
    \tag{P2}
\end{equation}

Given the positive-definiteness assumption on $\mathbf{\Sigma}$, \eqref{prob:Markowitz} is convex and thus $\mathbf{y}^\star$ exists.
Over time, several variants of \eqref{prob:Markowitz} have been proposed to consider modifications in the objective function, regularizations, and additional constraints (\cite{Perrin2020machine} and refs. therein).
Notably, the works of \cite{Puelz2001,Larsen2002,Gaivoronski2005,Fabozzi2010} represent variants considering \varr (and thus the P\&L distribution) as a measure of risk.
Furthermore, \cite{bertsimas1998optimal} consider the optimization of the vector of instrument notionals instead of $\mathbf{y}$. 
Hereinafter, we refer to the above variants of \eqref{prob:Markowitz} as M-approaches.

\paragraph{Distinguishing features}
As mentioned in \Cref{sec:introduction}, a first distinguishing feature of our approach is that we adopt the viewpoint of bank traders and risk managers, who focus on portfolio risk measurement and features at a given point in time $t$, and look for optimal trades at the same time $t$.
Mathematically, this means that unlike the M-approach, centered on forward-looking expected values of return and risk, in our approach the relevant portfolio risk measures and features are estimated based on historical scenarios of interest, as given in \Cref{sec:preliminaries}.

A second distinguishing feature between the proposed approach and the M-approach concerns the parameterization.
Specifically, as given in \Cref{eq:positionRATS}, our optimization variables are the positions of the particles, i.e., $2m$-dimensional integer vectors each entailing an optimization strategy \eligibleoptimizationstrategy, where $m \leq n$ is the desired cardinality of $\mathcal{H}$.
In contrast, in the M-approach, the variable to be optimized would be a vector of notionals in $\nat^{n}$, where $n$ is the cardinality of the UEI universe \UEIsuniverse.
As shown in the paragraph \enquote{Eligible optimization strategy} in \Cref{subsec:expset}, our parameterization allows the user to easily specify a (complex) structure for both \asstEOS and \notionalEOS.

A third distinguishing feature concerns the direct applicability to objective functions of the form in \Cref{eq:generalf}, i.e., general combinations of portfolio features and risk measures.
Indeed, the objective function in \eqref{prob:Markowitz} is convex and, although there are variants of \eqref{prob:Markowitz} that consider \varr (\cite{Fabozzi2010} and refs. therein), the fractional form of the objective function in \Cref{eq:obj_fun_cs} is a challenge for the M-approach.
In particular, the nonconvexity and potential non-differentiability of the objective function could impair the direct applicability of the M-approach to the RATPO problem.
Specifically, it would be necessary to investigate suitable convex and differentiable surrogates to tackle problems as \eqref{prob:optprob_cs}, as is done, for example, in the successive convex approximation optimization scheme extensively discussed by \cite{nedic2018parallel}.

Finally, another distinguishing feature concerns direct applicability in the presence of an arbitrarily complex set of constraints to be handled.
In fact, although sensitivity constraints in \Cref{eq:gconstraint_cs} are linear inequality constraints in the notional amounts (see \Cref{eq:Pfeat}), the group cardinality constraint entailed by the specified structure for \asstEOS would result in a constraint on the $\ell_0$-norm of groups of instruments associated with each underlying, where the groups must discriminate the type of instrument (see paragraph \enquote{Eligible optimization strategy} in \Cref{subsec:expset}).
Looking at the development of an M-approach for solving \eqref{prob:optprob_cs}, a possible way to enforce such a structure in the solution would be to replace the entailed group cardinality constraint with a convex regularization term within the objective function based on the mixed $\ell_{1}/\ell_{2}$-norm \footnote{Consider $\mathcal{I}=\{\mathcal{I}_1, \ldots, \mathcal{I}_u\}$ partition of $[n]$ and a vector $\mathbf{a} \in \reall^n$. Denote by $a_{\nu,i}$ the $i$-th entry within the subgroup $\nu$ of $\mathbf{a}$ entailed by $\mathcal{I}$. Then, the mixed $\ell_{p}/\ell_{q}$-norm of $\mathbf{a}$ is $$\norm{\mathbf{a}}_{p,q}\coloneqq \left(\sum_{\nu=1}^u \left( \sum_{i=1}^{\abs{\mathcal{I}_\nu}} \abs{a_{\nu,i}}^p \right)^{\frac{q}{p}} \right)^{\frac{1}{q}}\,.$$} of the vector of notional amounts.
Acting at the level of the groups, the $\ell_2$-norm would promote sparsity in the solution without causing the vanishing of the notional amounts of an entire group of instruments.
Acting within groups, the $\ell_1$-norm would result in the zeroing of the notional amounts of some UEIs belonging to each group.
However, such a solution would introduce a bias in estimating the notional amounts, and it may not be possible to satisfy exactly the group cardinality imposed through our approach.
Conversely, due to our parametrization (see \Cref{eq:positionRATS}), the proposed approach easily allows a desired composition to be enforced in the solution.

We believe that developing an M-approach to solve \eqref{prob:optprob_cs} is an intriguing research question to investigate, as highlighted by the above challenges to be addressed, thus not only interesting from the operational perspective. 
We, therefore, plan to tackle this research question in the future.

\subsection{Small-sized eligible optimization strategy} \label{subsec:toy_portfolio}
In this case, we consider a single underlying, the Euro Stoxx 50 Index (.STOXX50E).
Given the discussion in the paragraph \enquote{Eligible optimization strategy} of \Cref{subsec:expset}, since the underlying is a stock index, $\mathcal{H}$ has to be composed of only two vanilla options and one future written on the specified underlying.
Given the domains corresponding to the Euro Stoxx 50 Index features in \Cref{tab:ueis_features}, we have $n=54$ UEIs.
Additionally, we set the values of $\tau^{\Delta}$, $\tau^{\mathcal{V}}$, and $\tau^{\Gamma}$ equal to $\tau^g$.
We investigate three different sensitivity constraint settings, that is, $\tau^g \in \{0.1 ,0.5, 1.0\}$.
The latter values represent tight, medium, and loose sensitivity constraints.

\begin{table}[t]
    \centering
    \caption{$\expectedpl^\asstG$, $\varr^\asstG$, trading cost $c^\asstG$, and corresponding objective function value when $\generalportfolio$ is \emph{(i)} \initialportfolio, \emph{(ii)} \optimaltotalportfolio for the small-sized EOS case computed via brute-force, and \emph{(iii)} \estimatedoptimaltotalportfolio for the large-sized EOS case retrieved by \algorithmname, in the different settings determined by the value of $\tau^g$.
    Notice that \initialportfolio is the same in all case studies and settings.
    Additionally, since in the small-sized EOS case RATS retrieves a solution within the optimal solution set (cf. \Cref{fig:cs1_hyperparams}), we report only \optimaltotalportfolio in the table, as it coincides with the estimated one.}
    \sisetup{
        detect-mode,
        table-alignment    = right,
        round-mode              = places,
        table-format = 6.0
        } 
    \begin{tabularx}{\textwidth}{C{2cm}C{1.5cm}cS[round-precision=0,table-format=5.0]S[round-precision=0,table-format=6.0]S[round-precision=0,table-format=5.0]S[round-precision=4, table-format=1.4]}
    \toprule
    Portfolio \generalportfolio & Case Study & \text{$\tau^g$} & {\text{$\expectedpl^\asstG$ (EUR)}} & {\text{$\varr^\asstG$ (EUR)}} & {\text{$c^\asstG$ (EUR)}} & {\text{$\widetilde{f}\left(\generalportfolio\right)$}} \\
    \midrule
    \initialportfolio & All & All & 33918.410368 & -735749.548100 & 0.000000 & -0.045543\\
    \midrule
    \multirow{3}{1.5cm}{\centering \optimaltotalportfolio} & \multirow{3}{1.5cm}{\centering Small-sized EOS} & 0.10 & 31875.411748 & -590677.060624 & 905.199651 & -0.051658\\
     &  & 0.50 & 32727.265720 & -585979.430595 & 1035.278210 & -0.053290 \\
     &  & 1.00 & 32633.870786 & -582059.744372 & 1026.717219 & -0.053503 \\
    \midrule
    \multirow{3}{1.5cm}{\centering \estimatedoptimaltotalportfolio} & \multirow{3}{1.5cm}{\centering Large-sized EOS} & 0.10 & 59674.185596 & -233572.567452 & 6570.481115 & -0.219425\\
     &  & 0.50 & 62149.864322 & -230545.786158 & 7732.727877 & -0.226655 \\
     &  & 1.00 & 61081.637283 & -222681.764085 & 7955.081228 & -0.228569 \\
    \bottomrule
    \end{tabularx}
    \label{tab:numerical_results_cs}
\end{table}

Since, in this case, the cardinality of the solution space is of the order of $10^8$, we can compute the optimal value for the objective function in \Cref{eq:obj_fun_cs} via brute force.
Appendix \ref{app:ueis_STOXX50} provides the parameters representing one of the solutions in $\widetilde{\mathcal{E}}^\star$, for the three settings.
Additionally, \Cref{tab:numerical_results_cs} shows the EUR value of the $\expectedpl^\asstG$, $\varr^\asstG$, and trading cost $c^\asstG$ for \initialportfolio and for \optimaltotalportfolio, along with those of the corresponding objective function, in the three settings.

Even though the $\expectedpl^\asstG$ is slightly lower, we see that the \varr of \optimaltotalportfolio is consistently lower than that of \initialportfolio in all settings.
Regarding the cost, it is null in the case of \initialportfolio due to the absence of transactions.
In the case of \optimaltotalportfolio, the cost is contained and similar in the three settings.
Accordingly, the objective function evaluated at \optimaltotalportfolio is lower as well.

Looking across the settings $\tau^g$, the objective function evaluated at \optimaltotalportfolio improves as $\tau^g$ increases.
This trend is mainly driven by the reduction in the (cost-adjusted) $\varr^\asstG$. 
This result aligns with the fact that the sensitivity constraints loosen as $\tau^g$ increases.

\begin{figure}[ht]
    \centering
    \includegraphics[width=\textwidth]{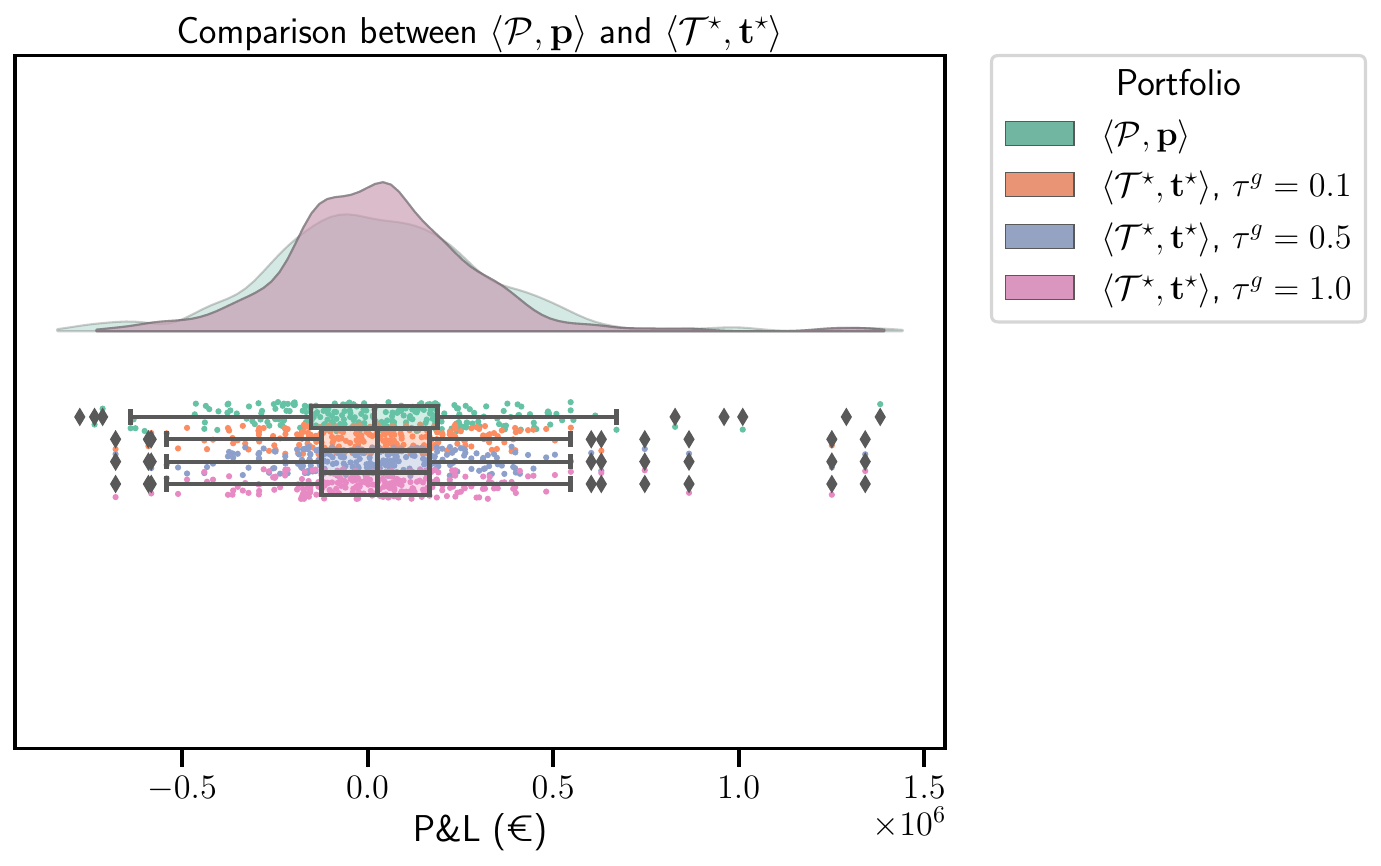}
    \caption{Empirical distribution of the P\&L of \initialportfolio and \optimaltotalportfolio, along with the risk scenarios P\&L represented as points and overlayed by the corresponding boxplot, for all the investigated settings determined by the value of $\tau^g$.}
    \label{fig:cs1_init_vs_best_total}
\end{figure}

Additionally, \Cref{fig:cs1_init_vs_best_total} depicts the empirical distribution of the P\&L of \initialportfolio and \optimaltotalportfolio, as well as the scenarios P\&L overlayed by the corresponding boxplot, for all the investigated settings.
Consistently with the provided cost-adjusted $\varr^\asstG$ reduction, in all the settings \optimaltotalportfolio pushes towards zero the left-tail of the distribution of \initialportfolio.
Furthermore, the empirical distributions associated with \optimaltotalportfolio are very similar in the three settings.

\begin{figure}[ht]
    \centering
    \includegraphics[width=\textwidth]{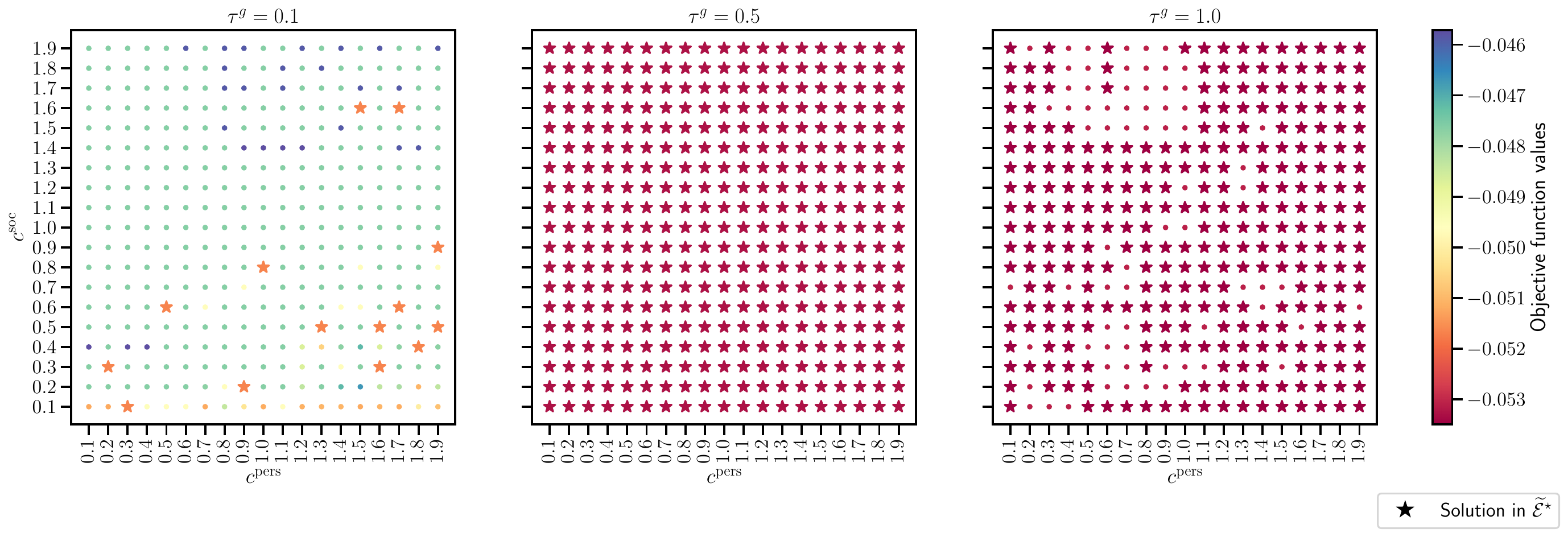}
    \caption{Objective function value corresponding to the solution \estimatedoptimaltotalportfolio estimated by \algorithmname, in all the settings and for all the tested $(c^{\mathrm{pers}}, c^{\mathrm{soc}})$ pairs.}
    \label{fig:cs1_hyperparams}
\end{figure}

At this point, to compare the solution retrieved by \algorithmname with the optimal one found via brute force, we solve \eqref{prob:optprob_cs} through \algorithmname using $n^p=1000$ particles, setting $k^{\mathrm{max}}=500$, and by investigating its behavior when both $c^{\mathrm{pers}}$ and $c^{\mathrm{soc}}$ take value in the discretized interval $(0., 2.)$, where we move with a step size equal to $0.1$.
The complete list of the \algorithmname hyper-parameters values is given in Appendix \ref{app:hyperparameters}.
We remark that, although the number of particles is large, the time required by \algorithmname to perform the allowed maximum number of iterations for solving \eqref{prob:optprob_cs} is only of the order of 10 seconds.
This holds also for the second application, see Appendix \ref{app:rt} for details.

The plots in \Cref{fig:cs1_hyperparams} depict the objective function value corresponding to the solution \estimatedoptimaltotalportfolio estimated by \algorithmname, in all the settings and for all the tested hyper-parameters values.
Within each plot, we represent with a star $\star$ the case in which our algorithm returns a solution belonging to the optimal solution set $\widetilde{\mathcal{E}}^\star$ of the corresponding RATPO problem, i.e., a solution equivalent to that found via brute force.

Our results empirically prove that \algorithmname can retrieve a solution in the optimal solution set in all the settings.
In case $\tau^g \in \{0.5, 1.0\}$, the optimal solution is returned for the most pairs of hyper-parameters value $(c^{\mathrm{pers}}, c^{\mathrm{soc}})$, suggesting that \algorithmname is also robust to the hyper-parameters choice when the sensitivity constraints loosen.

Looking at the objective function values related to the color legend in \Cref{fig:cs1_hyperparams}, we see that, in case $\tau^g=0.5$, we have no variation as we always retrieve a solution in the optimal solution set.
Regarding the case $\tau^g=1.0$, most solutions belong to the optimal solution set.
We also observe a few nearly optimal solutions, represented as points within the plot, characterized by a dark red color, which is almost indistinguishable from that of the stars.
Instead, in the more restrictive case $\tau^g=0.1$, a consistent color gradient shows up, thus signaling that the fine-tuning of the hyper-parameters value becomes more important when the constraints tighten.
Nevertheless, the minima of the objective function in the three settings are close in value.

\subsection{Large-sized eligible optimization strategy}\label{subsec:real_portfolio}
In this case, we consider $u=13$ underlyings.
Hence, given the domains in \Cref{tab:ueis_features}, we have $n=620$ UEIs.
In this case, the cardinality of the solution space is of the order of $10^{110}$. 
Thus, it is not feasible to compute the optimal objective function value and corresponding $\widetilde{\mathcal{E}}^\star$ via brute force.

We apply \algorithmname by using the same values for the hyper-parameters as in \Cref{subsec:toy_portfolio}, provided in Appendix \ref{app:hyperparameters}.
Appendix \ref{app:ueis_STOXX50} provides the parameters within \estimatedoptimaltotalportfolio corresponding to the .STOXX50E underlying to enhance the comparison between the two case studies.
In addition, \Cref{tab:numerical_results_cs} gives the EUR value of the $\expectedpl^\asstG$, $\varr^\asstG$, and trading cost $c^\asstG$ for \estimatedoptimaltotalportfolio, along with those of the corresponding objective function, in the three settings.
Overall, we see that \algorithmname consistently improves the $\expectedpl^\asstG$ while reducing the $\varr^\asstG$ of \initialportfolio. 
This is also true when we consider the trading cost, which is contained and similar in the three settings.
Accordingly, the objective function evaluated at \estimatedoptimaltotalportfolio is consistently lower than that evaluated at \initialportfolio.

Overall, looking across the settings $\tau^g$, the objective values associated with the retrieved solutions are very close, suggesting that \algorithmname is able to find a suitable EOS also in case of tighter sensitivity constraints. 
In detail, the objective function evaluated at \estimatedoptimaltotalportfolio slightly improves as $\tau^g$ increases.
The reduction of the $\varr^\asstG$ mainly drives this trend. 
This result is consistent with that in \Cref{subsec:toy_portfolio}.

\begin{figure}[h!]
    \centering
    \includegraphics[width=\textwidth]{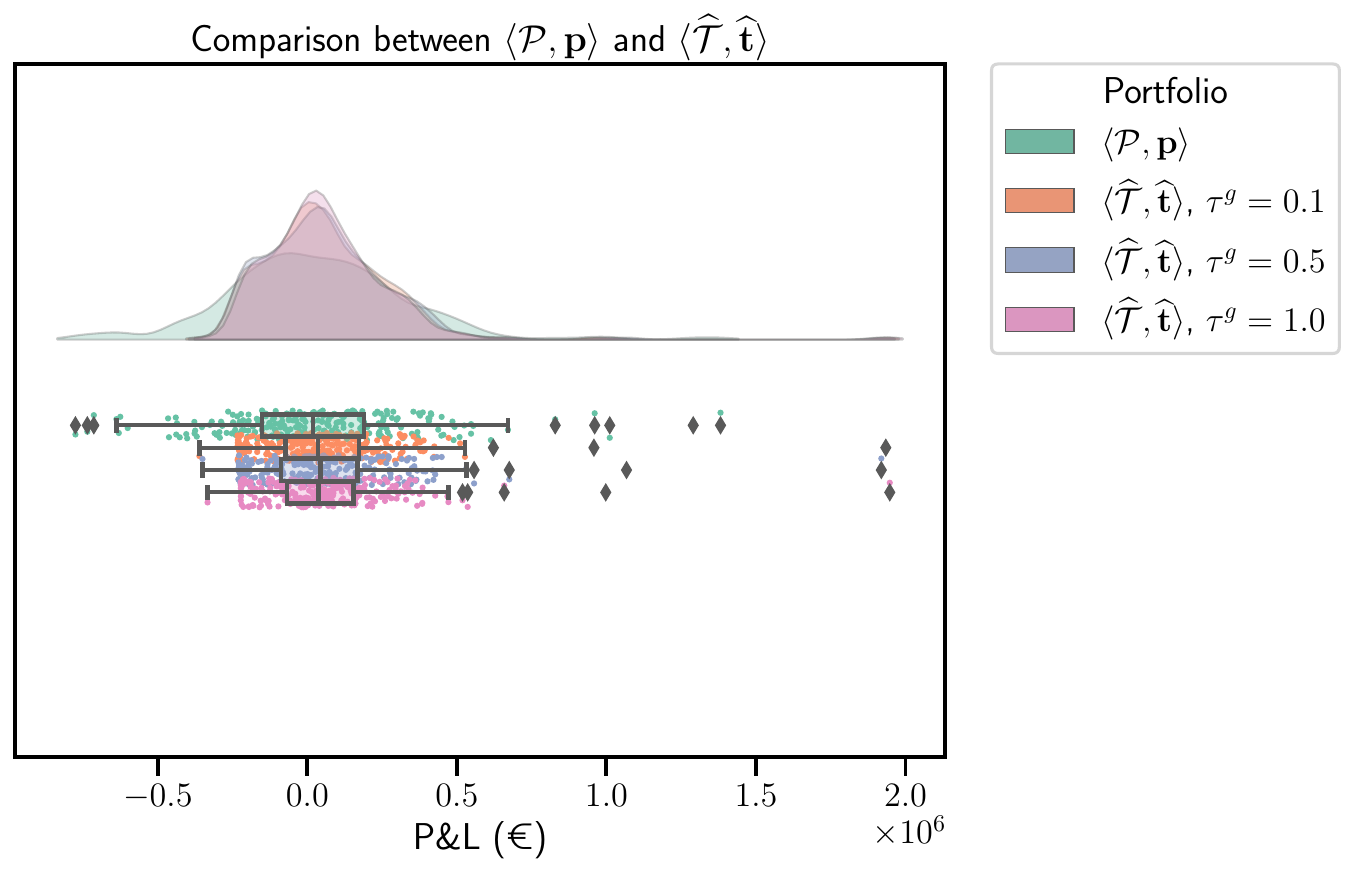}
    \caption{Empirical distribution of the P\&L of \initialportfolio and \estimatedoptimaltotalportfolio, along with the risk scenarios P\&L represented as points and overlayed by the corresponding boxplot, for all the investigated settings determined by the value of $\tau^g$.}
    \label{fig:cs2_init_vs_total}
\end{figure}

\Cref{fig:cs2_init_vs_total} depicts the empirical distribution of the P\&L of \initialportfolio and \estimatedoptimaltotalportfolio, as well as the scenarios P\&L overlayed by the corresponding boxplot, for all the investigated settings.
Consistently with the provided $\varr^\asstG$ reduction, in all the settings \estimatedoptimaltotalportfolio pushes towards zero the left-tail of the distribution of \initialportfolio.
Furthermore, the empirical distributions associated with \estimatedoptimaltotalportfolio are very similar in the three settings.
This further confirms that \algorithmname can find a performing EOS in all settings.

\begin{figure}[h!]
    \centering
    \includegraphics[width=\textwidth]{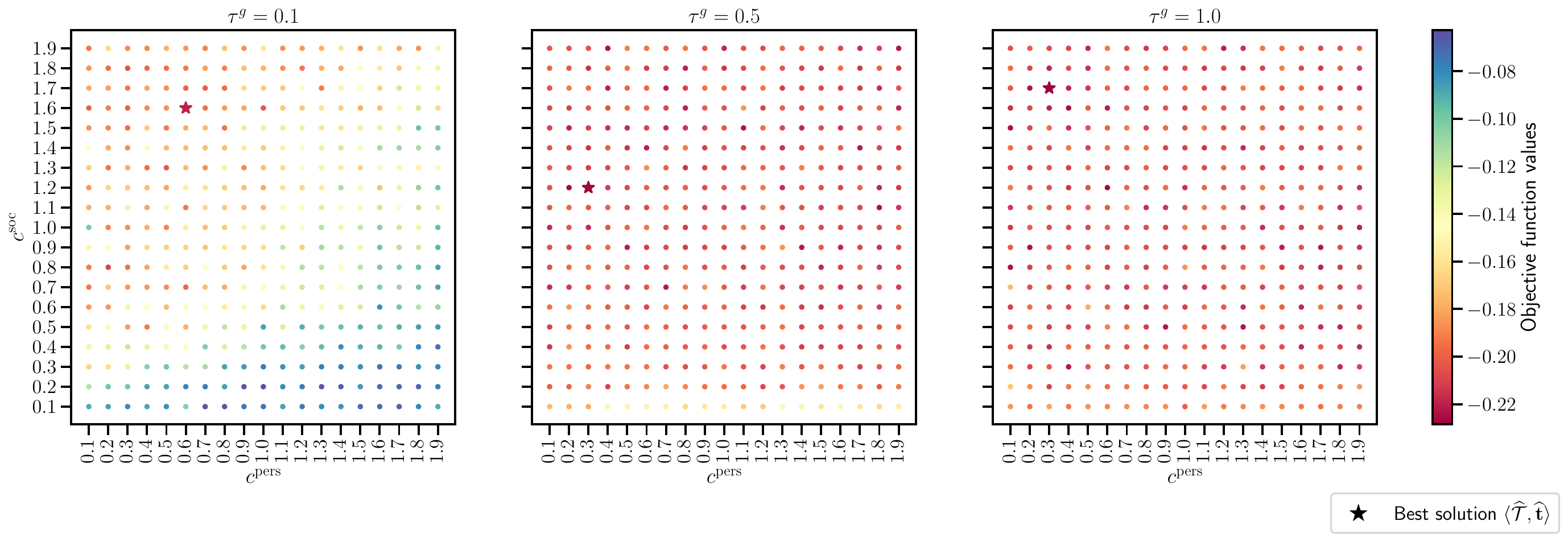}
    \caption{Objective function value corresponding to the solution \estimatedoptimaltotalportfolio estimated by \algorithmname, in all the settings and for all the tested $(c^{\mathrm{pers}}, c^{\mathrm{soc}})$ pairs.}
    \label{fig:cs2_hyperparams}
\end{figure}

Additionally, \Cref{fig:cs2_hyperparams} depicts objective function value corresponding to the solution \estimatedoptimaltotalportfolio estimated by \algorithmname, for the three settings and all the considered pairs $(c^{\mathrm{pers}}, c^{\mathrm{soc}})$.
In this case, for each setting, we put a star $\star$ in correspondence of the pair of values $(c^{\mathrm{pers}}, c^{\mathrm{soc}})$ that returns the best-estimated solution \estimatedoptimaltotalportfolio.
Looking at \Cref{fig:cs2_hyperparams}, in the case of looser sensitivity constraints ($\tau^g \in \{0.5, 1.0\}$), the objective function values corresponding to the retrieved solutions for the different hyper-parameters pairs become closer.
This confirms that \algorithmname is robust to the hyper-parameters choice when the sensitivity constraints loosen, as observed in \Cref{subsec:toy_portfolio}.
Additionally, in the case of tighter sensitivity constraints ($\tau^g=0.1$), the quality of the retrieved solution improves as we move toward the top-left corner of the grid.
This suggests that, from the particle viewpoint, having a stronger tilt toward the global best position at each iteration is beneficial to navigating the more challenging solution space.

\begin{figure}[h!]
    \centering
    \includegraphics[width=\textwidth]{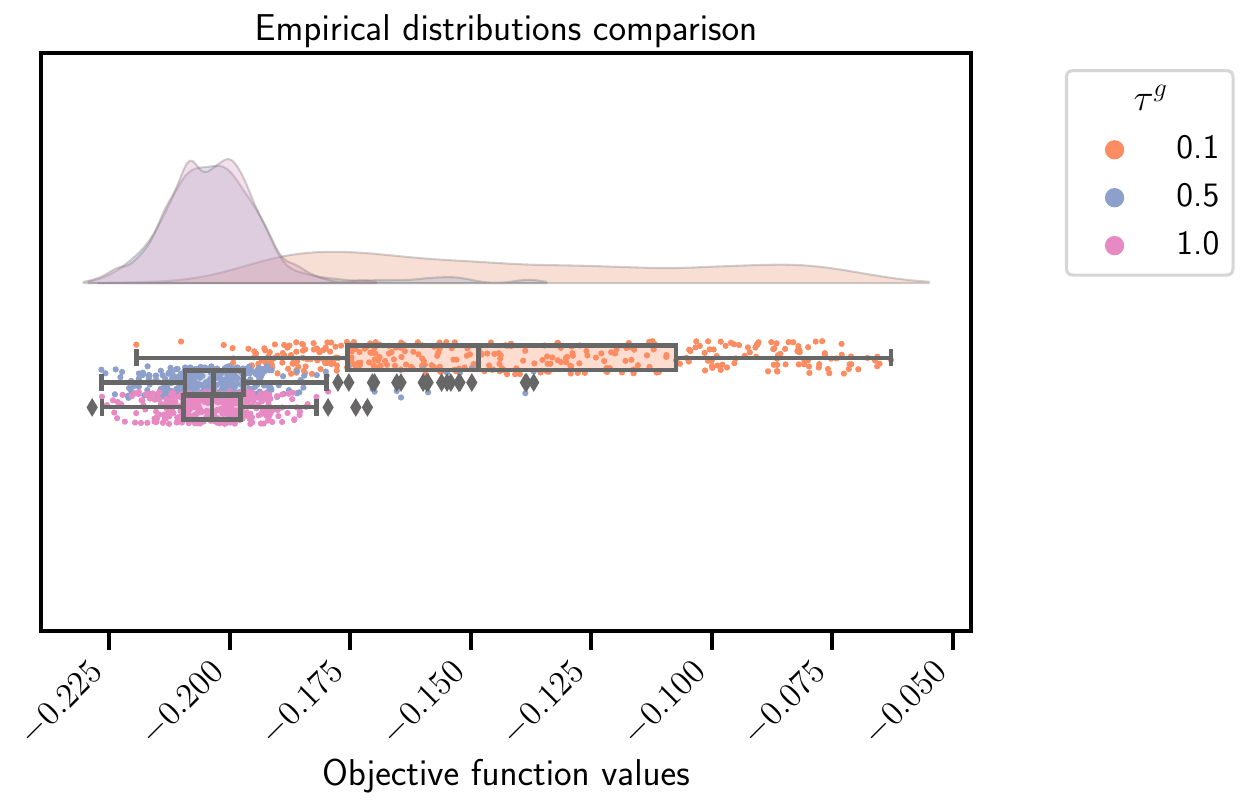}
    \caption{Empirical distributions of the objective function corresponding to the solution \estimatedoptimaltotalportfolio estimated by \algorithmname, in all the settings and for all the tested $(c^{\mathrm{pers}}, c^{\mathrm{soc}})$ pairs.}
    \label{fig:cs2_hyperparams_dist}
\end{figure}

Finally, \Cref{fig:cs2_hyperparams_dist} shows the empirical distributions of the objective function values in the three settings. 
Consistently with the discussion above, the distribution spread is lower in case $\tau^g \in \{0.5, 1.0\}$.
Nevertheless, the three minima are close in value.

\section{Conclusions and perspectives}\label{sec:conclusions}
We proposed and investigated the \emph{risk-aware trading portfolio optimization problem} (RATPO), 
which takes the point of view of a bank’s traders and risk managers, looking for optimal trades to reduce capital reserves while preserving portfolio value, consistently with the business objectives and limits. 
The problem solution leverages the basic financial concepts of unique eligible instruments and optimization strategies. 
 
We developed \algorithmname, a versatile computational tool that leverages the parallelizable nature of the optimization problem with respect to the particle dimension.
\algorithmname can be adapted to different trading portfolios, objective functions, and constraints.
Furthermore, \algorithmname may incorporate the financial insights and business views of traders and risk managers.

We provided a comprehensive numerical evaluation of the proposed approach through two real-world applications on trading portfolios. 
In the first application, we demonstrated the ability of \algorithmname to identify an optimal solution set, highlighting its advantages over M-approaches.\footnote{Recall that we use M-approach to refer to the variants of the quadratic programming problem \eqref{prob:Markowitz}.}
In the second application, we considered $13$ underlyings and $620$ unique eligible instruments, where the solution space grows to the order of $10^{110}$.
We demonstrated that \algorithmname was able to scale efficiently and handle the complexity of this larger problem, demonstrating its practical utility in large-scale portfolio optimization scenarios.

Our RATPO problem is general with respect to initial portfolios, risk measures and limits, eligible optimization instruments, trading strategies, and optimization algorithms. It can be applied to real portfolios of financial institutions. Strong financial insight is needed both to select the optimization parameters, i.e. the eligible optimization strategies, and to understand the financial soundness of the solutions proposed by \algorithmname.
In conclusion, our work bridges the gap between the implementation of effective trading strategies and compliance with stringent regulatory and economic capital requirements, allowing a better alignment of business and risk management objectives.
 
Future work, as discussed in~\Cref{subsec:M_approach}, could investigate the development of a gradient-based, continuous optimization M-approach to solve \eqref{prob:optprob_cs}. 
This research direction is relevant from an operational perspective and poses stimulating challenges. 
For example, a major issue is the non-convexity and potential non-differentiability of the objective function in \Cref{eq:obj_fun_cs}, which may hinder the direct application of these variants to portfolio optimization problems. 
Addressing this problem would involve the exploration of suitable convex and differentiable surrogates. 
Moreover, managing the complexities of the feasible set of eligible optimization strategies will be a critical area of our future research.
Indeed, the entailed constraints might include non-linear forms, and require methods to promote sparsity in the solutions without biasing the estimation of notional amounts.


\printbibliography[heading=bibintoc,title={References}]

\begin{appendices}

\section{Hyper-parameters}\label{app:hyperparameters}
Here we provide the detailed values of the hyper-parameters for the \algorithmname in Algorithm \eqref{algo:RATS} used in the empirical assessment in \Cref{sec:empirical_assessment}.

\begin{squishlist}
    \item Number of particles $n^p = 1000$;
    \item Particle personal coefficient $$c^{\mathrm{pers}} \in \{0.1, 0.2, 0.3, 0.4, 0.5, 0.6, 0.7, 0.8, 0.9, 1.0, 1.1, 1.2, 1.3, 1.4, 1.5, 1.6, 1.7, 1.8, 1.9\};$$
    \item Particle social coefficient $$c^{\mathrm{soc}} \in \{0.1, 0.2, 0.3, 0.4, 0.5, 0.6, 0.7, 0.8, 0.9, 1.0, 1.1, 1.2, 1.3, 1.4, 1.5, 1.6, 1.7, 1.8, 1.9\};$$
    \item Particle maximum velocity $v^{\mathrm{max}}=1.0$;
    \item Particle minimum velocity $v^{\mathrm{min}}=-1.0$;
    \item Particle maximum inertia $w^{\mathrm{max}}=1.0$;
    \item Particle minimum inertia $w^{\mathrm{min}}=1.0$;
    \item Significance threshold $\tau^f=10^{-4}$;
    \item Concentration threshold for the particle population $\tau^p=0.75$.
    \item Maximum number of iteration $k^{\mathrm{max}}=500$;
    \item Maximum number of of stall iteration $k^{\mathrm{max \,stall}}=100$.
\end{squishlist}

\section{Data for UEIs}\label{app:ueis_features_sn}

\begin{sidewaystable}[hbt]
    \centering
    \caption{The table shows \emph{(i)} the considered underlyings divided by category, i.e., stock and stock index; \emph{(ii)} the parameters to determine for each underlying $\ell$ the set of UEIs relevant for the EOS, that is, $\omega_{\ell,i}$, $K_{\ell,i}$, $T_{\ell,i}\,$; and \emph{(iii)} the ranges of the notional associated with the first two instruments (vanilla options) and the third instrument (stock/futures). Specifically, $21$ evenly-spaced values have been selected within the specified range (extremes and zero included).}
    \begin{tabular}{clrrrrr}
    \toprule
     & & $\omega_{\ell,i}$ & $K_{\ell,i}$ & $T_{\ell,i}$ & $[\ell_{\ell,i},\, t_{\ell,i}], \, i \in [2]$ & $[\ell_{\ell,i},\, t_{\ell,i}], \, i=3$\\
    Category & Underlying &  &  &  &  &  \\
    \midrule
    \multirow[t]{8}{*}{Stock} & $\nu=1:$ AXAF.PA & c, p, s & 0.10, 0.25, 0.50 & 21, 49, 84, 168, 266, 630 & [-760\,000, 760\,000] & [-340\,000, 340\,000] \\
     & $2:$ FB.O & c, p, s & 0.10, 0.25, 0.50 & 21, 49, 112, 168, 266, 476, 630 & [-120\,000, 120\,000] & [-50\,000, 50\,000] \\
     & $3:$ GTO.AS & c, p, s & 0.10, 0.25, 0.50 & 21, 49, 84, 168, 266, 630 & [-310\,000, 310\,000] & [-160\,000, 160\,000] \\
     & $4:$ IBM.N & c, p, s & 0.10, 0.25, 0.50 & 21, 49, 112, 168, 266, 476, 630 & [-130\,000, 130\,000] & [-60\,000, 60\,000] \\
     & $5:$ KO.N & c, p, s & 0.10, 0.25, 0.50 & 21, 49, 112, 168, 266, 476, 630 & [-410\,000, 410\,000] & [-190\,000, 190\,000] \\
     & $6:$ ORCL.N & c, p, s & 0.10, 0.25, 0.50 & 21, 49, 84, 168, 266, 476, 630 & [-370\,000, 370\,000] & [-140\,000, 140\,000] \\
     & $7:$ PG.N & c, p, s & 0.10, 0.25, 0.50 & 21, 49, 112, 168, 266, 476, 630 & [-230\,000, 230\,000] & [-110\,000, 110\,000] \\
     & $8:$ T.N & c, p, s & 0.10, 0.25, 0.50 & 21, 49, 112, 168, 266, 476, 630 & [-570\,000, 570\,000] & [-280\,000, 280\,000] \\
    \midrule
    \multirow[t]{5}{*}{Stock Index} & $9:$ .FTMIB & c, p, q & 0.10, 0.25, 0.50 & 21, 49, 84, 168, 266, 630 & [-800, 800] & [-500, 500] \\
     & $10:$ .FTSE & c, p, q & 0.10, 0.25, 0.50 & 21, 49, 84, 168, 266, 630 & [-2\,000, 2\,000] & [-2\,000, 2\,000] \\
     & $11:$ .GSPC & c, p, q & 0.10, 0.25, 0.50 & 21, 49, 84, 168, 266, 476, 630 & [-7\,000, 7\,000] & [-3\,000, 3\,000] \\
     & $12:$ .NDX & c, p, q & 0.10, 0.25, 0.50 & 21, 49, 84, 168, 266, 476, 630 & [-3\,000, 3\,000] & [-1\,000, 1\,000] \\
     & $13:$ .STOXX50E & c, p, q & 0.10, 0.25, 0.50 & 21, 49, 84, 168, 266, 630 & [-5\,000, 5\,000] & [-3\,000, 3\,000]\\
    \bottomrule
    \end{tabular}
    \label{tab:ueis_features}
\end{sidewaystable}

\Cref{tab:ueis_features} shows the considered underlyings for the empirical assessment in \Cref{sec:empirical_assessment}, as well as the corresponding parameters to determine the UEIs according to \Cref{def:UEI_application}.

The ranges of the notional amounts for the options were selected to cover the Vega sensitivity of the initial portfolio, namely $\mathcal{V}^{\mathcal{P}}$.
The ranges for stocks and futures were selected to cover the Delta sensitivity of the initial portfolio, namely $\Delta^{\mathcal{P}}$.
Specifically, $\forall \, \ell \in [u]$, indicate with $\Delta^{\mathrm{c}}_{\ell}$ and $\Delta^{\mathrm{p}}_{\ell}$ the Delta sensitivity of the at-the-money (ATM) call and put options with the highest maturity associated with the $\ell$-th underlying. 
Analogously, $\forall \, \ell \in [u]$, indicate with $\mathcal{V}^{\mathrm{c}}_{\ell}$ the Vega sensitivity of the ATM call option with the highest maturity associated with the underlying.
Hence, define 
\begin{equation}
    \eta_{\ell}^a \coloneqq \frac{\mathcal{V}^{\mathcal{P}}}{\abs{\mathcal{V}^{\mathrm{c}}_{\ell}}}\,, \quad \text{ and } \quad \eta_{\ell}^b \coloneqq \frac{\Delta^{\mathcal{P}}}{\mathrm{max}(\abs{\Delta^{\mathrm{c}}_{\ell}}, \abs{\Delta^{\mathrm{p}}_{\ell}})}\,. 
\end{equation}

At this point, round both $\eta_{\ell}^a$ and $\eta_{\ell}^b$ to the closest hundred, thousand, and so on, depending on the order of magnitude.
Mathematically, given $x \in \reall$, the applied rounding reads as
\begin{equation}\label{eq:rounding}
    \bar{x} = \left\lceil \frac{1}{2} \left\lfloor 2 \frac{x}{10^{\left\lfloor \log_{10}(x)\right\rfloor}} \right\rfloor \right\rceil \cdot 10^{\left\lfloor\log_{10}(x) \right\rfloor} .
\end{equation}

For instance, according to \Cref{eq:rounding}, for $x=75$ we have $\bar{x}=80$; for $x=740$ we have $\bar{x}=700$.  
Hence, $\forall \, \ell \in [u]$, the ranges of notional amounts are given by

\begin{equation}\label{eq:notionals_range}
    [\ell_{\ell,i},\, t_{\ell,i}] = \begin{cases}
         [-\bar{\eta}_{\ell}^a, \bar{\eta}_{\ell}^a], & \mathrm{if} \; i \in \{1,2\};\\
         [-\bar{\eta}_{\ell}^b, \bar{\eta}_{\ell}^b], & \mathrm{if} \; i=3.
    \end{cases}
\end{equation}

Starting from \Cref{eq:notionals_range}, for each $i \in [n^i]$ and $\ell \in [u]$, the domain of the notional amount $\mathcal{D}^{h_{\ell,i}}$ is given by $21$ evenly-spaced integer values (including $0$) in $[\ell_{\ell,i},\, t_{\ell,i}]$.

\clearpage
\section{Running time}\label{app:rt}

This appendix provides details regarding the total running time required by \algorithmname for solving \eqref{prob:optprob_cs}.
Since in the two applications given in \Cref{subsec:toy_portfolio,subsec:real_portfolio} the iterations run by \algorithmname might vary over different simulations (each determined by a specific pair of values $(c^{\mathrm{pers}}, c^{\mathrm{soc}})$), here we monitor the total running time (measured in seconds) divided by the performed number of iterations.
We remark that this is not exactly the time per iteration, as the total running time also includes the cost of the initialization (cf. Algorithm \ref{algo:RATS}).

\begin{figure}[h!]
    \centering
    \includegraphics[width=0.8\linewidth]{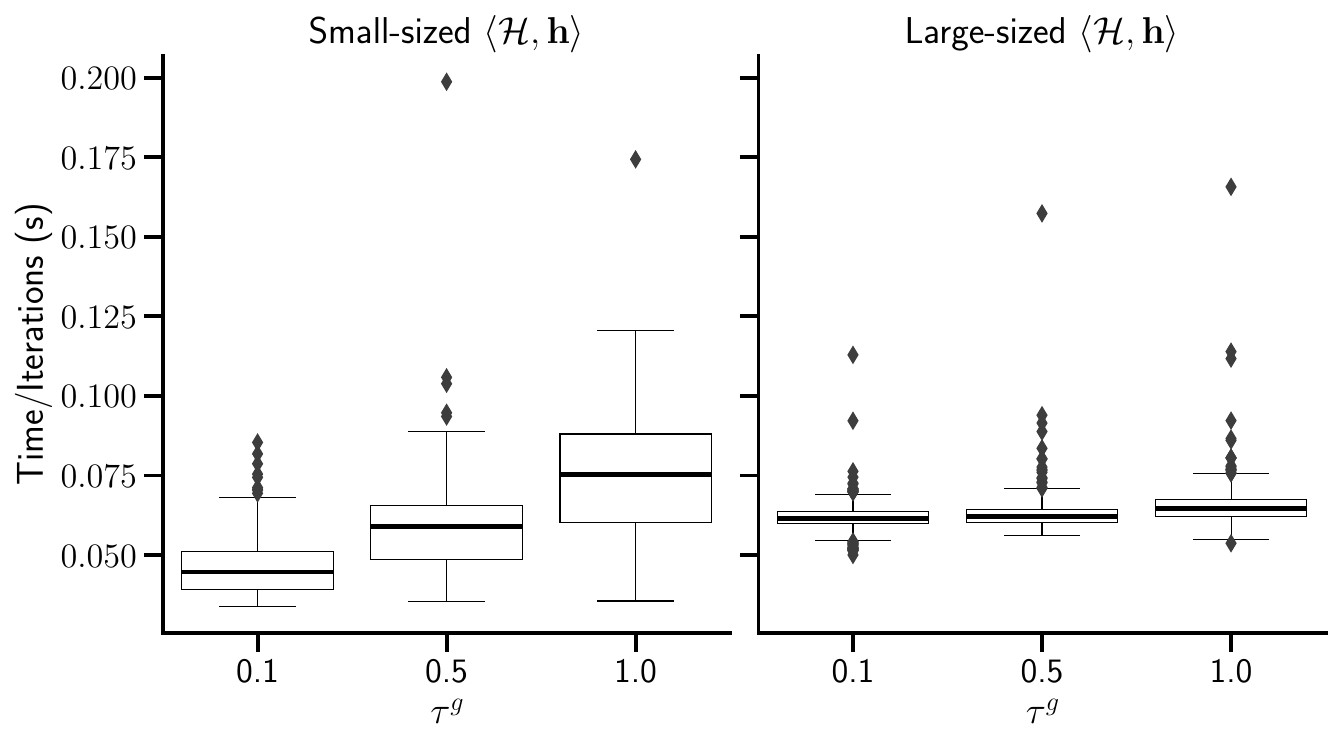}
    \caption{Box plot for the total running time (measured in seconds) divided by the performed number of iterations for (left) the first and (right) second application, and for all the considered values of $\tau^g$.
    For each application and for each value of $\tau^g$, we create the box plot with the running time and number of iterations associated with the $361$ simulations corresponding to the tested pairs $(c^{\mathrm{pers}}, c^{\mathrm{soc}})$.}
    \label{fig:rt}
\end{figure}

\Cref{fig:rt} provides the box plot corresponding to the monitored quantity for (left) the first and (right) second application, and for all the considered values of $\tau^g$.
For each application and for each value of $\tau^g$, we create the box plot with the running time and number of iterations associated with the $361$ simulations corresponding to the tested pairs $(c^{\mathrm{pers}}, c^{\mathrm{soc}})$.
Given the hyper-parameters in Appendix \ref{app:hyperparameters}, from the figure we deduce that the time required by \algorithmname for running $k^{\mathrm{max}}$ iterations is roughly of the order of $10$ seconds, in both case studies.
This empirically proves that performing \algorithmname takes a short time, even when the number of particles is large.

Even though it is beyond the scope of this appendix, from \Cref{fig:rt} we notice that in the first application, the monitored quantity seems to increase along $\tau^g$, while in the second, it shows no dependence.
We hypothesize that one of the causes of this behavior is a side-effect of the usage of just-in-time (JIT) compilation in our code.

\begin{figure}[h!]
    \centering
    \begin{subfigure}[b]{\textwidth}
        \centering
        \includegraphics[width=\textwidth]{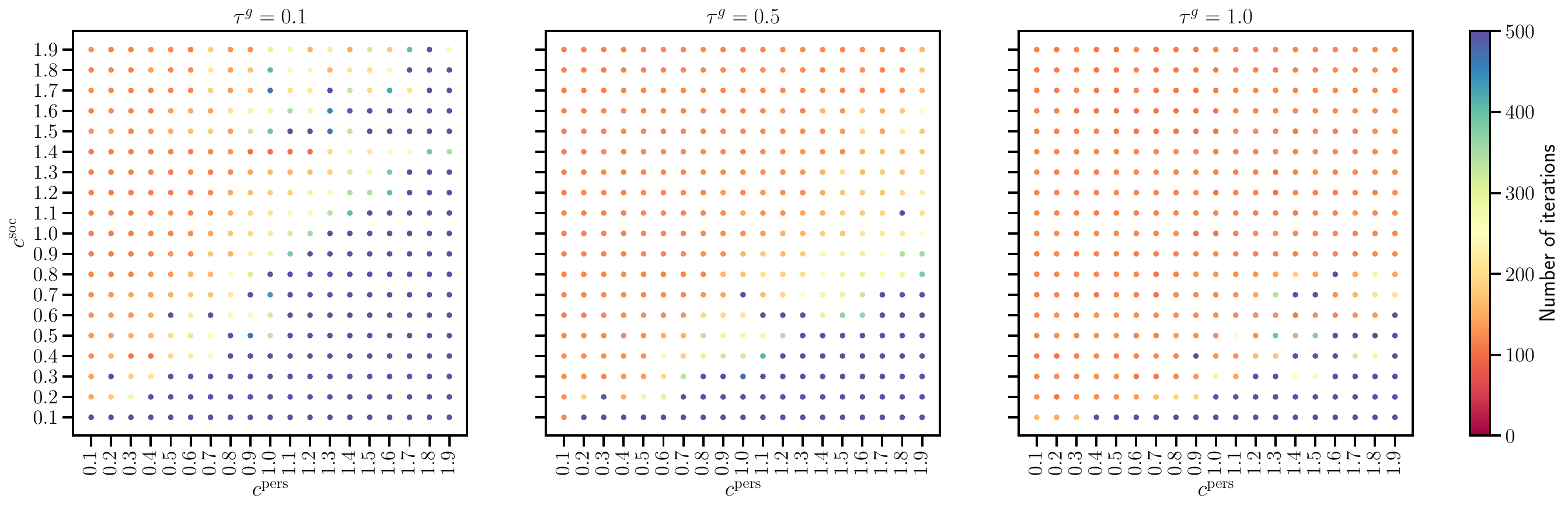}
        \caption{Small-sized EOS}
        \label{subfig:niter_cs1}
    \end{subfigure}
    
    \vspace{1em} 

    \begin{subfigure}[b]{\textwidth}
        \centering
        \includegraphics[width=\textwidth]{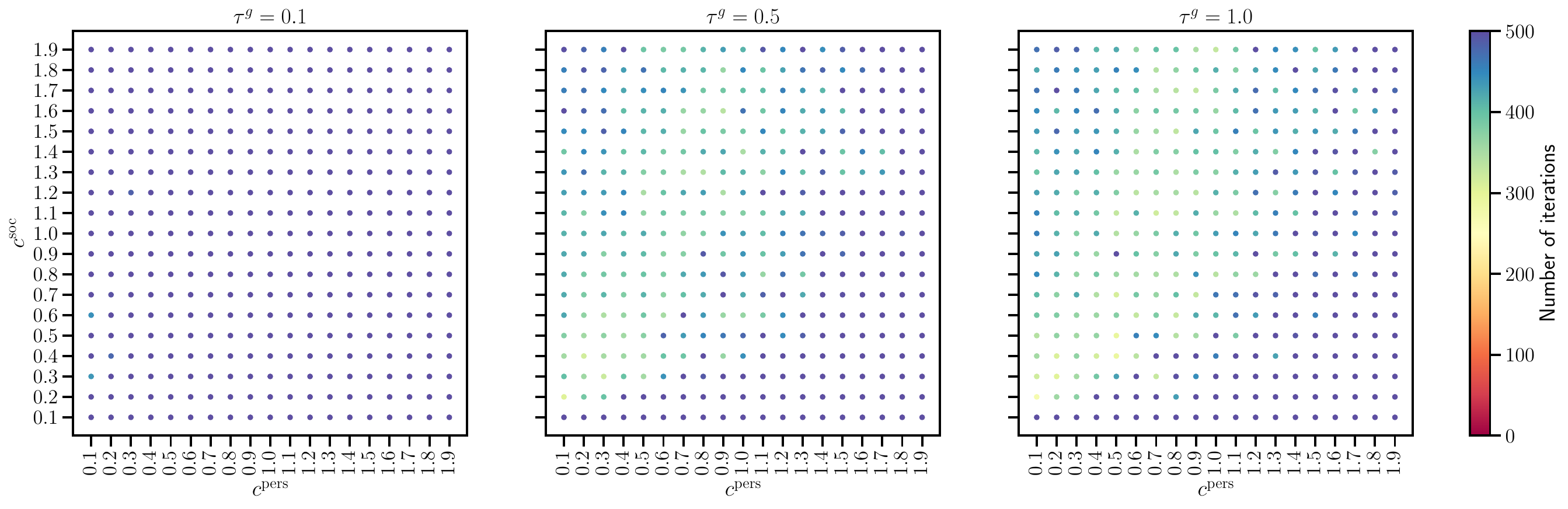}
        \caption{Large-sized EOS}
        \label{subfig:niter_cs2}
    \end{subfigure}

    \caption{Number of iterations for the tested $(c^{\mathrm{pers}}, c^{\mathrm{soc}})$ for (\subref{subfig:niter_cs1}) the first and (\subref{subfig:niter_cs2}) second application.}
    \label{fig:niter}
\end{figure}

When using JIT compilation, there is an initial loading cost due to compiling code at runtime. 
This cost affects the total execution time, and its impact is larger when the number of iterations that reuse the precompiled code is low.

\Cref{fig:niter} shows the number of iterations for the tested $(c^{\mathrm{pers}}, c^{\mathrm{soc}})$ for (\subref{subfig:niter_cs1}) the first and (\subref{subfig:niter_cs2}) second application.
From \Cref{subfig:niter_cs1} we see that as $\tau^g$ increases, the number of simulations performing a number of iterations lower than $k^{\mathrm{max}}$ increases.
Hence, the impact of the loading cost increases as well.
This explains the (false) dependence on $\tau^g$ of the value of the total execution time divided by the number of iterations observed in \Cref{subfig:niter_cs1}. 
Conversely, \Cref{subfig:niter_cs2} shows that the number of iterations performed by \algorithmname is roughly the same in all the investigated settings.
Accordingly, in this case, we do not observe any dependence in \Cref{subfig:niter_cs2}.

\begin{figure}[h!]
    \centering
    \begin{subfigure}[b]{\textwidth}
        \centering
        \includegraphics[width=\textwidth]{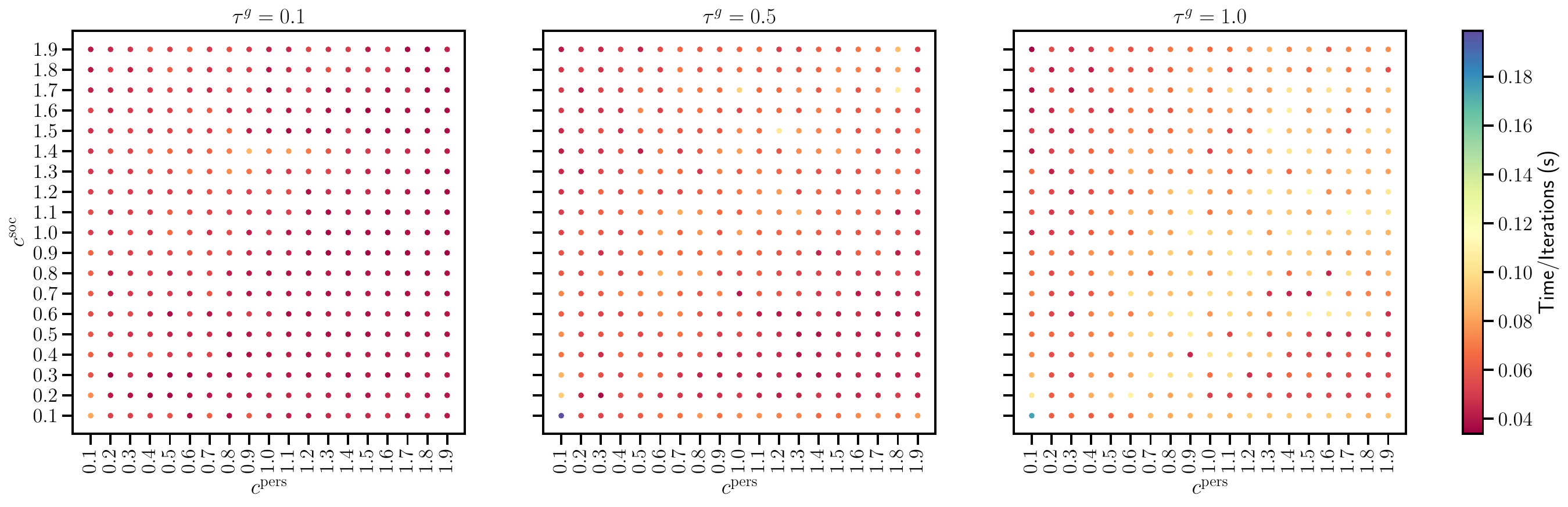}
        \caption{Small-sized EOS}
        \label{subfig:ti_cs1}
    \end{subfigure}
    
    \vspace{1em} 

    \begin{subfigure}[b]{\textwidth}
        \centering
        \includegraphics[width=\textwidth]{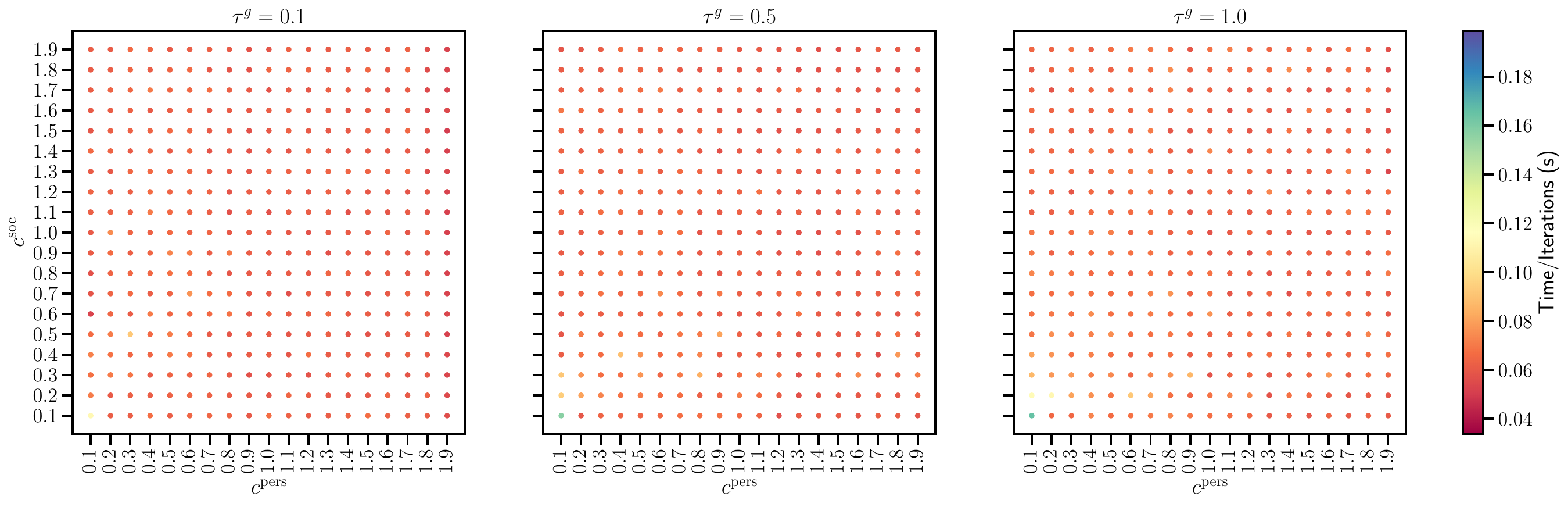}
        \caption{Large-sized EOS}
        \label{subfig:ti_cs2}
    \end{subfigure}

    \caption{Total running time (measured in seconds) divided by the performed number of iterations for (\subref{subfig:ti_cs1}) the first and (\subref{subfig:ti_cs2}) second application, and for all the considered values of $\tau^g$.}
    \label{fig:ti}
\end{figure}

This side-effect is further visualized in \Cref{fig:ti}, which shows that the monitored value tends to be higher for the simulations that are shown to perform a fewer number of iterations for the optimization of the EOS in \Cref{fig:niter}. 

\section{Estimated parameters for Euro Stoxx 50 Index}\label{app:ueis_STOXX50}

\begin{table}[!h]
    \centering
    \caption{Parameters $\omega_{\ell, i}$, $K_{\ell, i}$, $T_{\ell, i}$, and $h_{\ell, i}$ identifying the selected UEIs and corresponding notionals for the .STOXX50E underlying within the solution retrieved by \algorithmname.
    We report the parameters' values for all the settings investigated in the empirical assessment provided in \Cref{sec:empirical_assessment}.}
    \sisetup{
        detect-mode,
        table-alignment    = right,
        round-mode              = places,
        round-precision = 2
        } 
    \begin{tabular}{C{3.5cm}C{2.5cm}SrrrS[round-precision=0,table-format=3.0]S[round-precision=0,table-format=4.0]}
        \toprule
        Case Study & Underlying & {\text{$\tau^g$}} & $i$ & $\omega_{\ell,i}$ & $K_{\ell,i}$ & {\text{$T_{\ell,i}$}} & {\text{$h_{\ell,i}$}} \\
        \midrule
        \multirow{9}{3.5cm}{\centering Small-sized EOS} & \multirow{9}{2.5cm}{\centering .STOXX50E} & {\multirow{3}{*}{0.10}} & 1 & c & 0.50 & 049 & -3500.000000 \\
         &  &  & 2 & p & 0.25 & 021 & 5000.000000 \\
         &  &  & 3 & q & 0.10 & 021 & 3000.000000 \\
        \cmidrule{3-8}
         &  & {\multirow{3}{*}{0.50}} & 1 & c & 0.50 & 266 & -4500.000000 \\
         &  &  & 2 & p & 0.10 & 049 & 5000.000000 \\
         &  &  & 3 & q & 0.10 & 021 & 3000.000000 \\
        \cmidrule{3-8}
         &  & {\multirow{3}{*}{1.00}} & 1 & c & 0.50 & 266 & -5000.000000 \\
         &  &  & 2 & p & 0.10 & 021 & 5000.000000 \\
         &  &  & 3 & q & 0.10 & 021 & 3000.000000 \\
        \midrule
        \multirow{9}{3.5cm}{\centering Large-sized EOS} & \multirow{9}{2.5cm}{\centering .STOXX50E} & {\multirow{3}{*}{0.10}} & 1 & c & 0.25 & 630 & 0.000000 \\
         &  &  & 2 & c & 0.50 & 168 & -4000.000000 \\
         &  &  & 3 & q & 0.25 & 021 & -600.000000 \\
        \cmidrule{3-8}
         &  & {\multirow{3}{*}{0.50}} & 1 & c & 0.10 & 084 & 0.000000 \\
         &  &  & 2 & c & 0.50 & 630 & -4000.000000 \\
         &  &  & 3 & q & 0.25 & 021 & 600.000000 \\
        \cmidrule{3-8}
         &  & {\multirow{3}{*}{1.00}} & 1 & c & 0.50 & 266 & -3000.000000 \\
         &  &  & 2 & p & 0.10 & 021 & 5000.000000 \\
         &  &  & 3 & q & 0.25 & 021 & 900.000000 \\
        \bottomrule
    \end{tabular}
    \label{tab:solutions_STOXX50}
\end{table}

\Cref{tab:solutions_STOXX50} provides, as an example, the solutions retrieved by \algorithmname for both case studies described in \Cref{sec:empirical_assessment}, for all the investigated settings. 
As a reminder, in the small-sized EOS case, the solutions estimated by \algorithmname are equivalent to the optimal obtained through brute force.
It can be observed that \algorithmname selects varying quantities of call/put options and futures (in the large-sized EOS case, only the .STOXX50E underlying is shown for simplicity), adhering to their respective discrete domains (see \Cref{tab:ueis_features}) and the imposed constraints. 
Notably, in the large-sized EOS case study, the two null notional amount values in the rightmost column indicate instances where \algorithmname identifies simplified trading strategies as more effective, involving only one option rather than two.
This behavior is not explicitly favored in our implementation.

\end{appendices}

\end{document}